\pdfoutput=1
\RequirePackage{ifpdf}
\ifpdf % We~are running pdfTeX in pdf mode
\documentclass[pdftex]{sigma}
\else
\documentclass{sigma}
\fi

\usepackage{mathrsfs}
\usepackage{dsfont}

\numberwithin{equation}{section}

\newtheorem*{Theorem*}{Theorem}

 { \theoremstyle{definition}

 }

\newcommand{\be}{\begin{equation}}
\newcommand{\ee}{\end{equation}}
\newcommand{\ba}{\begin{aligned}}
\newcommand{\ea}{\end{aligned}}
\newcommand{\ben}{\begin{eqnarray}\displaystyle}
\newcommand{\een}{\end{eqnarray}}

\newcommand{\p}{\partial}

\def\({\big(}
\def\){\big)}
\def\[{\left[}
\def\]{\right]}
\newcommand{\CA}{{\cal A}}

\newcommand{\CF}{{\cal F}}

\newcommand{\CH}{{\cal H}}

\newcommand{\CK}{{\cal K}}
\newcommand{\CL}{{\cal L}}
\newcommand{\CM}{{\cal M}}

\newcommand{\CO}{{\cal O}}

\newcommand{\CS}{{\cal S}}
\newcommand{\CT}{{\cal T}}
\newcommand{\CU}{{\cal U}}

\newcommand{\CX}{{\cal X}}

\newcommand{\CZ}{{\cal Z}}
\newcommand{\cA}{{\cal A}}

\newcommand{\cO}{{\cal O}}

\def\IZ{{\mathbb Z}}
\def\IR{{\mathbb R}}
\def\IC{{\mathbb C}}
\def\IN{{\mathbb N}}
\def\IP{{\mathbb P}}

\newcommand{\re}{{\rm e}}
\newcommand{\ri}{{\rm i}}
\newcommand{\rd}{{\rm d}}

\newcommand{\oD}{\mathsf{D}}
\newcommand{\mO}{\mathsf{O}}

\newcommand{\mW}{\mathsf{W}}

\def\ty{{\tilde{y}}}

\def\tUps{\tilde\Upsilon}
\def\bt{\boldsymbol{t}}

\newcommand{\cX}{{\cal X}}
\def\tzeta{\tilde\zeta}
\def\I{{\rm i}}
\def\eps{\epsilon}
\newcommand{\talp}{\tilde\alpha}
\newcommand{\teta}{\tilde\eta}

\newcommand{\txi}{\tilde\xi}
\newcommand{\Li}{{\rm Li}}
\def\Ab{\mathbf{A}}

\def\sign{{\rm sgn}}
\def\by{\bar y}

\def\bft{\boldsymbol{t}}
\def\bfu{\boldsymbol{u}}
\def\bfn{\boldsymbol{n}}
\def\bfc{\boldsymbol{c}}
\def\bfd{\boldsymbol{d}}
\def\bfz{{\boldsymbol{z}}}

\def\snl{s_\ell}
\def\snk{s_k}

\def\Heis{\mathds{H}}

\def\scT{\mathscr{T}}

\def\scV{\mathscr{V}}
\def\scU{\mathscr{U}}

\newcommand{\mb}{{\mathsf{b}}}

\renewcommand{\Re}{\operatorname{Re}}
\renewcommand{\Im}{\operatorname{Im}}

\newcommand{\fad}{\operatorname{\Phi}_{\mathsf{b}}}
\newcommand{\fadone}{\operatorname{\Phi}_{1}}

\begin{document}

%\allowdisplaybreaks

\newcommand{\arXivNumber}{2311.17638}

\renewcommand{\PaperNumber}{073}

\FirstPageHeading

\ShortArticleName{Resurgence of Refined Topological Strings and Dual Partition Functions}

\ArticleName{Resurgence of Refined Topological Strings\\ and Dual Partition Functions}

\Author{Sergey ALEXANDROV~$^{\rm a}$, Marcos MARI\~NO~$^{\rm b}$ and Boris PIOLINE~$^{\rm c}$}

\AuthorNameForHeading{S.~Alexandrov, M.~Mari\~no and B.~Pioline}

\Address{$^{\rm a)}$~Laboratoire Charles Coulomb (L2C), Universit\'e de Montpellier, CNRS, \\
\hphantom{$^{\rm a)}$}~F-34095, Montpellier, France} % Address of First Author
\EmailD{\href{mailto:sergey.alexandrov@umontpellier.fr}{sergey.alexandrov@umontpellier.fr}}

\Address{$^{\rm b)}$~D\'epartement de Physique Th\'eorique et Section de Math\'ematiques,\\
\hphantom{$^{\rm b)}$}~Universit\'e de Gen\`eve, Gen\`eve, CH-1211 Switzerland}
\EmailD{\href{mailto:Marcos.Marino@unige.ch}{Marcos.Marino@unige.ch}}

\Address{$^{\rm c)}$~Laboratoire de Physique Th\'eorique et Hautes Energies (LPTHE), UMR 7589,\\
\hphantom{$^{\rm c)}$}~CNRS-Sorbonne Universit\'e, Campus Pierre et Marie Curie, 4 place Jussieu, \\
\hphantom{$^{\rm c)}$}~F-75005 Paris, France}
\EmailD{\href{mailto:pioline@lpthe.jussieu.fr}{pioline@lpthe.jussieu.fr}}

\ArticleDates{Received December 13, 2023, in final form August 02, 2024; Published online August 06, 2024}

\Abstract{We study the resurgent structure of the refined topological string partition function on a non-compact Calabi--Yau threefold, at large orders in the string coupling constant $g_s$ and fixed refinement parameter $\mathsf{b}$. For $\mathsf{b}\neq 1$, the Borel transform admits two families of simple poles,
corresponding to integral periods rescaled by $\mb$ and $1/\mathsf{b}$. We show that the corresponding Stokes automorphism is expressed in terms of a~generalization of the non-compact quantum dilogarithm, and we conjecture that the Stokes constants are determined by the refined Donaldson--Thomas invariants counting spin-$j$ BPS states. This jump in the refined topological string partition function is a special case (unit five-brane charge) of a more general transformation property of wave functions on quantum twisted tori introduced in earlier work by two of the authors. We show that this property follows from the transformation of a suitable refined dual partition function across BPS rays, defined by extending the Moyal star product to the realm of contact geometry.}

\Keywords{resurgence; topological string theory; Borel resummation; Stokes automorphism}

\Classification{81T45; 14N35; 34M40; 34M55}

\section{Introduction and discussion}

As a mathematically well-defined subsector of type II superstring theories, topological string theory
provides a prime arena for exploring the non-perturbative completion of the asymptotic series
predicted by the perturbative genus expansion in string theory. In cases where the target
Calabi--Yau threefold $X$ admits a large $N$ dual, a non-perturbative formulation is available,
which resums the perturbative series
to all orders in the topological string coupling $g_s=1/N$, for fixed values of the moduli $\bft$
(corresponding to 't Hooft couplings in the dual gauge theory)
at least for integer values of $N$.
Non-perturbative corrections of order ${\rm e}^{-1/g_s}$ typically arise from tunneling (or instanton)
effects in the dual matrix model.
In general however, one has only access to the genus-$g$ free energy $\CF_g(\bft)$ occurring
at order $g_s^{2g-2}$ in the perturbative expansion, and non-perturbative corrections are ambiguous,
without further assumptions.

Assuming that the putative,
non-perturbative topological string partition function $\CF(\bft;g_s)$ belongs to a suitable
class of resurgent functions, one can however extract much information about non-perturbative corrections
from the growth of the coefficients $\CF_g(\bft)$ at large genus. This method was proposed early on in \cite{Shenker:1990uf}
for general string theory models, and first applied to the topological string in \cite{mmnp,mm-open,msw,ps09}.
In particular, the instanton action $\cA$
controlling the size of the leading instanton effects ${\rm e}^{-\cA/g_s}$ can be read off from the location
of the singularity closest to the origin in the Borel plane.

While the explicit computation of the free energies $\CF_g(\bft)$ is usually impractical at large genus,
they are strongly constrained by the holomorphic anomaly equations (HAE) \cite{bcov}.
Following~\mbox{\cite{cesv2,cesv1}}, much progress has been made recently in constructing the trans-series
solution to the HAE \cite{Gu:2023mgf,gm-multi,Iwaki:2023cek,ms}. In particular, it was shown that the instanton
actions $\cA$ entering in the trans-series are equal (up to overall normalization) to the central
charge $Z(\gamma)$ of vectors $\gamma$ in the
charge lattice (the Grothendieck lattice $K_0(X)$ in the $A$-model, or the
homology lattice $H_3(X,\IZ)$ in the B-model).
Moreover, the Stokes automorphism
$\mathfrak{S}$
controlling the jump in the Borel resummation
when a radial integration contour crosses a sequence of singularities was determined in \cite{Iwaki:2023cek}:
the jump induces the following transformation in the partition function~${\CZ={\rm e}^{\CF}}$,
\be
\label{SZ}
\CZ\ \mapsto\ \mathfrak{S} \CZ =
\exp\[\frac{ \omega(\gamma)}{2\pi \ri}\big(\Li_2\big({\rm e}^{2\pi x}\big)+2\pi x\log\big(1-{\rm e}^{2\pi x}\big)\big)\] \CZ,
\ee
where $\omega(\gamma)$ is (up to normalization) the so-called Stokes constant,
\begin{gather*}
\Li_2(z)=\sum_{k\geq 1} z^k/k^2
\end{gather*}
 is the dilogarithm function and $x$ is suitable operator
\big(equal to $\frac{\ri g_s}{2\pi} c^a \partial_{t^a}$ in the simplest case considered in \cite{Iwaki:2023cek},
which we also restrict to here for simplicity; here $c^a$ are the components of $\gamma$ in
$H_4(X,\IZ)$, interpreted as D4-brane charges in type IIA set-up, or D3-brane charges in
type IIB\big). As observed in \cite{Iwaki:2023cek}, the transformation
\eqref{SZ} is exactly such that the so-called dual partition function
\be
\label{deftau}
\tau( \bft, \bfu; g_s) = \sum_{\bfn\in\IZ^n} {\rm e}^{2\pi \bfn \cdot\bfu/g_s} \CZ(\bft-\ri g_s\bfn; g_s)
\ee
transforms by a simple factor, up to a $\bfu$-dependent shift in $\bft$,
\be
\label{Stau}
\mathfrak{S} \tau( \bft, \bfu;g_s) ={\rm e}^{\frac{\omega(\gamma)}{2\pi\I} \Li_2(\re^{2 \pi {\bfc} \cdot \bfu/g_s})}
\tau \left( \bft+ \frac{\omega(\gamma)}{2\pi} g_s {\bfc} \log\(1-\re^{2 \pi {\bfc} \cdot \bfu/g_s}\) , \bfu; g_s \right).
\ee
We note that dual partition functions (which are obtained as so-called Zak transforms of the conventional partition
functions) have appeared in many different
contexts, including supersymmetric gauge theories \cite{no2},
topological string theory \cite{adkmv,dhsv},
topological recursion \cite{Eynard:2021sxg,em},
the spectral theory/topological string correspondence \cite{ghm}
and isomonodromic tau functions~\cite{CLT,cpt,GIL-12,GIL-13}.

Quite remarkably, the transformations \eqref{SZ} and \eqref{Stau} have appeared before in yet a different context,
namely the study of five-brane instanton corrections to the hypermultiplet
moduli space in type II strings compactified on a Calabi--Yau threefold \cite{APP,Alexandrov:2015xir}.
The key idea of~\smash{\cite{APP,Alexandrov:2015xir}} is that
S-duality relates NS5-brane instanton corrections in type IIB string theory to D5-brane instanton corrections,
which (for unit D5-brane charge) are governed by the topological string partition function
by virtue of \cite{gw-dt,gw-dt2}. More specifically, for fixed unit D5-brane charge ${k=1}$, the sum over
D3-D1-D(-1) brane charges leads to a non-Gaussian theta series of the form \eqref{deftau} (with~$\bfn$ being the D3-brane charge). Due to wall-crossing behavior of the Donaldson--Thomas (DT)
invariants $\Omega(\gamma)$ counting D5-D3-D1-D(-1) instantons in type IIB
(or equivalently D6-D4-D2-D0 black holes in type IIA, for in either case, these invariants
count semi-stable objects in the derived category of coherent sheaves on $X$),
the theta series becomes
a section of a~certain line bundle~$\CL$ over the (twistor space of) the hypermultiplet moduli space,
with gluing conditions given by~\eqref{Stau}, which in turn imply the transformation~\eqref{SZ}
for the kernel of the theta series.
This suggests in particular that the Stokes constant $\omega(\gamma)$ should be equated with the DT
invariant~$\Omega(\gamma)$ counting BPS states associated to the same ray, as proposed
in different contexts in \cite{astt, Gu:2023wum, Gu:2023mgf,gm-peacock,gm-multi,mm-s2019}.
These considerations were extended in \cite{Alexandrov:2015xir} in two directions: first,
to $k>1$ units of five-brane charge, where the theta series becomes a section of the $k$-th
power of the line bundle $\CL$, and second, to include the effect of the refinement parameter $y$
in Donaldson--Thomas theory, conjugate to angular momentum, which induces a non-commutative deformation
of the (twistor space of the) hypermultiplet moduli space.

This brings us to the main goal of the present paper, which is to extend the results
\eqref{SZ}--\eqref{Stau} to the refined topological string partition function.
The latter can be viewed as a one-parameter deformation $\CF(\bft;g_s,\mb)$ of the
usual topological free energy $\CF(\bft;g_s)$, which exists on non-compact
Calabi--Yau threefolds with a $\IC^\times$ action.\footnote{It was suggested in
\cite{hkatzk} that the refined topological string could also be defined on compact CY threefolds,
but the resulting free energies are no longer independent of complex structure moduli,
and it is unclear if the HAE still holds.}
While the worldsheet definition of refined topological strings for $\mb\neq 1$ remains obscure
(see \cite{Antoniadis:2013bja} for an early attempt),
in space-time it is interpreted as the partition function of the five-dimensional
gauge theory engineered by M-theory compactified on $X$ times a
five-dimensional $\Omega$-background \cite{n,Nekrasov:2010ka} with parameters
\be
\label{ep-beta-intr}
\epsilon_1= g_s \mb,
\qquad
\epsilon_2 = -g_s \mb^{-1}.
\ee
The resulting partition function leads to integer BPS invariants $N_{\bfd}^{j_L,j_R}$
which refine the usual
Gopakumar--Vafa and Donaldson--Thomas invariants \cite{ckk,hiv,no}.
When $X$ is a toric CY singularity, $\CF(\bft;g_s,\mb)$ can be computed
by a refined version \cite{ikv} of the topological vertex formalism~\cite{akmv}.
Otherwise, the main computational tool is the refined version \cite{hk} of the
standard holomorphic anomaly equations (HAE) \cite{bcov}, which can be integrated
inductively with sufficient control on the behavior at boundaries of moduli space.

In this work, we extend the analysis of \cite{gm-multi,Iwaki:2023cek} and construct the general
trans-series solution to the refined HAE.\footnote{The trans-series solution in the NS limit
of the refined theory \cite{ns} requires a particular treatment and
was studied in \cite{gm-qp}.} In particular, we find that there are now two instanton
actions~$\mb Z(\gamma)$ and~$\mb^{-1} Z(\gamma)$ contributing to the trans-series
for each $\gamma$ in the charge lattice.\footnote{The occurrence of two instanton actions $\cA/\mb$ and $\mb\cA$
was already observed
in an earlier study of the resurgent structure of the refined topological string on the resolved
conifold \cite{Grassi:2022zuk}. Our main claim is that this feature arises for refined topological
strings on arbitrary non-compact CY threefolds.}
Moreover, in the simplest case where the singularity
corresponds to a BPS state with unit Stokes constant~${\omega(\gamma)=1}$ (as befits
states which become massless at a conifold singularity), we find that the Stokes automorphism
\eqref{SZ} is deformed to
\be
\label{SZb}
\mathfrak{S}_{\mb} \CZ = \Phi_{\mb}^{-1} (x)  \CZ,
\ee
where $ \Phi_{\mb}(x)$ is the Faddeev quantum dilogarithm defined in \eqref{defPhib}, reducing
to \eqref{SZ} at $\mb=1$ due to \eqref{onePhi}.
More generally, the Stokes automorphism associated to a singularity supported
at~$Z(\gamma)$ is given by
\be
\label{SZbj}
\mathfrak{S}_{\mb} \CZ = \prod_{j} \Bigg[
\prod_{m=-j}^j \Phi_\mb\big(x+\ri m\big(\mb-\mb^{-1}\big)\big)
 \Bigg]^{-\omega_{[j]}} \CZ,
\ee
where the product runs over half integer $m$ such that $m+j$ is integer, and $\omega_{[j]}$ are Stokes constants.
This has a natural interpretation in terms of the motivic DT invariants $\Omega(\gamma,y)$
introduced in \cite{KS:stability} (see also \cite{Dimofte:2009bv}). The latter can be decomposed as
\begin{gather*}
\Omega(\gamma,y)=\sum_j \Omega_{[j]}(\gamma) \chi_j(y),
\end{gather*}
 where
\be
\chi_{j} (y)
:= \sum_{m=-j}^j y^{2m} =
\frac{y^{2j+1}-y^{-2j-1}}{y-y^{-1}}
\label{defcharacter}
\ee
is the character of the spin $j$ representation of ${\rm SU}(2)$.
We conjecture that the Stokes constants~$\omega_{[j]}$ associated to the singularity at $Z(\gamma)$ are equal to the coefficients
$\Omega_{[j]}(\gamma)$ appearing in this decomposition.\footnote{The relation between the refinement parameters $\mb$ and $y$ will become apparent in \eqref{defyb}.}
This conjecture is motivated by the comparison
of \eqref{SZbj} with a structurally identical formula obtained in \cite{Alexandrov:2015xir} in a different but related context, as we will
review and extend in Section \ref{refined-sec}.
The conjecture can be also verified directly in special cases, e.g.,
when~$\gamma$ corresponds to a D2-D0 BPS state, as shown below in equation~\eqref{trans-LR}.

It is then
natural to ask whether there exists a deformation of the dual partition function \eqref{deftau},
with kernel given by the refined topological string partition function $\CZ$, whose transformation rules
across rays would be guaranteed by the Stokes automorphism \eqref{SZb}, or more generally~\eqref{SZbj}.
In fact, the Stokes automorphism \eqref{SZb} is a special case $k=1$ of an operator
\smash{${\bf A}_{\mb^2}^{(k)}$} acting on $L^2(\IR)\otimes \IC^k$ which was introduced in \cite[Section~4]{Alexandrov:2015xir}
in a rather {\it ad hoc} fashion, for the purpose of producing a doubly-quantized version
of the standard 5-term relation for the quantum dilogarithm. In this work, we
provide a rationale for the construction of \cite[Section~4]{Alexandrov:2015xir}, by explicitly constructing a
generalized theta series valued in a degree $k$ line bundle $\CL_{\mb}$ over the quantum torus,
which is well defined across BPS rays provided the kernel of the theta series transforms according
to \smash{${\bf A}_{\mb^2}^{(k)}$}, or more generally by a $j$-dependent extension of \smash{${\bf A}_{\mb^2}^{(k)}$}.
For $k=1$, this provides a refined version of the dual partition function \eqref{deftau}.
Both the theta series and its transformation under wall-crossing are defined using a certain
Moyal-type non-commutative product on twistor space.
In fact, the non-commutativity makes it possible to introduce two different refined dual partition functions
which are both related to the refined topological string partition function, but with one of
the twistor space coordinates identified with (the inverse of)
either of the two deformation parameters $\eps_{1,2}$ from \eqref{ep-beta-intr}.

These results raise several obvious questions: first, can one prove the equality between
Stokes constants and Donaldson--Thomas invariants, perhaps in the framework of the Riemann--Hilbert
problem proposed in \cite{Alexandrov:2021wxu,Barbieri:2019yya,Bri-19,Bri-23,Bridgeland:2020zjh}?
Second, what is the physical interpretation of the refined dual partition function,
for example in terms of free fermions, spectral determinants, tau functions,
partition functions of line defects, or otherwise?
Third, does there exists a \mbox{rank-$k$} version of the topological string
(refined or unrefined) whose large order behavior would be governed by
the Stokes automorphism \smash{${\bf A}_{\mb^2}^{(k)}$}, and which
would be related to the partition function of~$k$ five-branes constructed
in \cite[Section~5]{APP} by a rank-$k$ version of the GV/DT correspondence~\cite{gw-dt}?
We hope to return to some of these challenging problems in the near future.

The remainder of this work is organized as follows. In Section~\ref{sec_reftop}, we review basic properties
of the refined topological string free energy. In Section~\ref{sec_trans}, we construct the trans-series solution
of the holomorphic anomaly equations, determine the boundary conditions at
the conifold locus and in the large volume limit, deduce the Stokes automorphism associated
to the singularities in the Borel plane, and perform numerical checks in the case of $X=K_{\IP^2}$.
In Section~\ref{refined-sec}, we recover the same automorphism
and generalize it to any integer $k\geq 1$ by constructing a class of dual partition functions on quantum twisted tori.
Definitions and properties of several variants of the quantum dilogarithm function,
which plays a central role in this work, are collected in Appendix~\ref{sec_qdilog}.

\section{The refined topological string}
\label{sec_reftop}

In this section, we review some
basic facts about the refined topological string and how to calculate its free energy.

As mentioned above, the refined topological string is parametrized by two complex
parameters~$\epsilon_{1,2}$. Setting
\be
\label{ep-beta}
\epsilon_1= g_s \mb, \qquad \epsilon_2 = -g_s \mb^{-1},
\ee
the parameter $g_s$ can be identified as the topological string coupling constant, while $\mb$ can
be regarded as a deformation parameter.
When $\mb=1$ one recovers the standard topological string. We note that many references (like \cite{kw})
use a parameter $\beta$, which is related to ours simply by~${\mb= \beta^{1/2}}$.
The refined topological string free energy is a formal power series \cite{hk}
\be
\CF({\bft}; \epsilon_1,\epsilon_2) =
\sum_{i,j\geq 0} (\epsilon_1+\epsilon_2)^{2i}  (-\epsilon_1\epsilon_2)^{j-1}\CF^{(i,j)}(\bft),
\ee
where the coefficients $\CF^{(i,j)}(\bft)$ are sections of an appropriate line bundle
on the moduli space of~$X$ (of K\"ahler structures in the A-model, or complex structures in the B-model),
parametrized by the flat coordinates ${\bft}$. (These are complexified K\"ahler parameters in the A-model,
and periods of the holomorphic 3-form in the B-model. We recall that the coordinates are flat with respect to the Gauss--Manin connection.)
For the purposes of this paper, we regard the free energy as a formal power series in $g_s$ whose coefficients
depend on the moduli and on the deformation parameter~$\mb$,
\be
\CF({\bft}; g_s,\mb)= \sum_{g \ge 0} g_s^{2g-2}  \CF_g({\bft}; \mb),
\ee
where
\be
\label{refined-fgb}
\CF_g({\bft}; \mb)= \sum_{k=0}^g \CF^{(k, g-k)}({\bft})  \big(\mb- \mb^{-1}\big)^{2k}.
\ee
The deformed free energies $\CF_g({\bft}; \mb)$ are Laurent polynomials in $\mb$, of degree $2g$,
and invariant under the exchange $\mb \leftrightarrow \mb^{-1}$. When $\mb=1$,
 \be
 \CF_g({\bft}; 1)= \CF^{(0,g)}({\bft})=\CF_g({\bft})
 \ee
is the conventional, unrefined topological string free energy at genus $g$.
The amplitudes $\CF^{(k, g\!-\!k)}\!({\bft})$ are related to the ones defined in \cite{hk} by an overall sign $(-1)^{k}$.

The refined free energies $\CF_g({\bft}; \mb)$ can be calculated on certain local CY geometries
using instanton calculus~\cite{n}, and more generally
with the refined topological vertex of~\cite{ikv}. These A-model-like
calculations are intrinsically attached to the large radius of the geometry.
In the refined case,
the B-model approach to the free energies is based on the HAE of~\cite{bcov}.\footnote{In the
unrefined case,
and in local CY geometries, the B-model is described by the topological recursion of~\cite{eo}, as proposed in \cite{bkmp,mm-open}.
To our knowledge a refined version of the formalism
for generic local CY geometries is not available yet, see
however \cite{Kidwai:2022kxx} for recent progress.}
An extension of the HAE to the refined case was proposed in~\cite{hk}
(see also \cite{kw}), and tested in detail in, e.g., \cite{ckk,hkk}.
In the framework of the HAE, one extends
the free energies $\CF_g({\bft}; \mb)$ to non-holomorphic functions, and the HAE control their non-holomorphic dependence.
We will use Roman capital letters for the non-holomorphic free energies obtained from the HAE,
and curly capital letters for their holomorphic limit,
as in \cite{Gu:2023mgf,gm-multi}.

To write the HAE, we need some ingredients from special geometry (see, e.g., \cite{klemm} for more
details). In the local case, the mirror CY is encoded in an algebraic curve
usually called the mirror curve. We recall that the
moduli space $\CM$ of complex structures of the mirror CY is a~special
K\"ahler manifold of complex dimension $n$. We will denote
by $z^a$, with $a=1,\dots, n$ a~set of algebraic complex coordinates on this moduli space.
The K\"ahler metric (here $\partial_a=\partial_{z^a}$)
\begin{gather}
G_{a \bar b} =\partial_a \partial_{\bar b} K
\end{gather}
derives from a K\"ahler potential $K$, which is determined
from the prepotential $\CF_0$, equal to the
genus zero free energy (the latter does not depend on $\mb$, as it clear from \eqref{refined-fgb}).
We introduce the covariant
derivative $D_a$ for the Levi--Civita connection associated to the metric,
\be
\Gamma^a_{bc} = G^{a \bar k} \partial_b G_{c \bar k}.
\ee
In the case of compact CY threefolds, the covariant derivative involves as well a
connection on the Hodge bundle $\CL$ over $\CM$. However, in the local case,
this connection vanishes in the holomorphic limit. Therefore, we can solve the HAE by setting formally
$K_a=0$ \cite{kz}. An additional ingredient we will need is the Yukawa coupling, which is a
tensor $C_{abc}$. In the coordinate system given by the flat coordinates $t^i$, the Yukawa coupling
is given by the third derivatives of
the genus 0 free energy, so we have in general
\be
C_{abc} = {\partial t^i \over \partial z^a} {\partial t^j \over \partial z^b} {\partial t^k \over \partial z^c} {\partial^3 \CF_0
\over\partial t^i \partial t^j \partial t^k}.
\ee
Finally, we have
to introduce an anti-holomorphic version of the Yukawa coupling, defined by%
\be
\label{antic}
{\overline C}_{\bar c}^{a b}= \re^{2 K}  G^{a \bar d} G^{b \bar e}{\overline C}_{\bar a \bar d\bar e } .
\ee
We can now write the refined HAE, following \cite{hk}. In terms of the free energies $F^{(k,g-k)}$, they read
\be
\partial_{\bar c} F^{(k, g-k)}= {1\over 2} \overline C^{ab}_{\bar c} \bigg( D_a D_b F^{(k, g-k-1)} +
\sum_{m, h} D_a F^{(m,h)} D_b F^{(k-m, g-k-h)} \bigg),
\ee
where $g\ge 2$ and the sum over $m$, $h$ is such that $m+h \ge 1$.
However, it is easy to check that, in terms of the combinations $F_g(\mb)$ defined in \eqref{refined-fgb},
the HAE take the same form as in the unrefined case of \cite{bcov}:
\be
\label{rhae-nf}
\partial_{\bar c} F_g (\mb)= {1\over 2} \overline C^{ab}_{\bar c} \left( D_a D_b F_{g-1}(\mb)
+ \sum_{h=1}^{g-1} D_a F_h (\mb) D_b F_{g-h}(\mb) \right), \qquad g\ge 2.
\ee
(Here, we only indicate the dependence of the refined free energies on the deformation parameter,
but they of course also depend on $z^a$.) Therefore, the refined free energies satisfy
the same HAE as the unrefined ones, but the starting point of the recursion, given by the
free energies at $g=1$, is different and given by
\be
\label{full-g1}
F_1(\mb)= F_1 (\mb=1)+ \big(\mb- \mb^{-1}\big)^2 F^{(1,0)}.
\ee
Here, $F_1 (\mb=1)=F^{(0,1)}$ is the conventional, unrefined free energy at genus one.
The free energy~$F^{(1,0)}$ turns out to be given by a holomorphic function of the moduli. It has the form~\mbox{\cite{hkk,hk}}
\begin{gather}
\label{f1ns}
F^{(1,0)}= -{1\over 24} \log ( f(\bfz) \Delta(\bfz) ),
\end{gather}
where $\Delta(\bfz)$ is the discriminant of the mirror curve, and $f(\bfz)$ is a rational function of the moduli,
explicitly known in many cases. The fact that $F^{(1,0)}$ is holomorphic will be important later on.

As in the unrefined local case, it is useful to introduce propagators $S^{ab}$, defined by the condition that \cite{bcov}
\be
\partial_{\bar c} S^{ab}= \overline C^{ab}_{\bar c}.
\ee
As a result, the HAE \eqref{rhae-nf} can be rewritten in the form
\be
\label{rhae-prop}
{\partial F_g (\mb) \over \partial S^{ab}}= {1\over 2} \left( D_a D_b F_{g-1}(\mb)
+ \sum_{h=1}^{g-1} D_a F_h (\mb) D_b F_{g-h}(\mb) \right), \qquad g\ge 2.
\ee
The initial condition \eqref{full-g1} has to be also re-expressed in terms of propagators.
Since $F^{(1,0)}$ is purely holomorphic, it does not
depend on the propagators, whereas for $F^{(0,1)}$ one has \cite{bcov}
\be
\label{F1prop}
\partial_a F^{(0,1)}= {1\over 2} C_{abc} S^{bc} + f_a(\bfz).
\ee
In this equation, $f_a(\bfz)$ is a function of the
moduli only. The equation \eqref{rhae-prop}, together with the
initial conditions \eqref{full-g1}, \eqref{F1prop} and \eqref{f1ns},
provides a recursive procedure to obtain the free energies $F_g(\mb)$ as polynomials
in the propagators, which is sometimes called ``direct integration" \cite{al,gkmw,hk06,yy}. At each
genus $g$ one has to fix an integration constant,
independent of the propagators, but dependent on the moduli, and called the holomorphic ambiguity
(the functions~$f_a(\bfz)$ in \eqref{F1prop}
can be regarded as examples thereof).

We would like to recall that the holomorphic free energies $\CF_g(\bt; \mb)$ depend on what is called
a~choice of {\it symplectic frame}. This choice
is determined, in the B-model, by a choice of a~symplectic basis of periods, which determines in turn
a choice of flat coordinates $\bt$ and of ``dual" coordinates $\partial \CF_0/\partial \bt$. A symplectic transformation of the periods
leads to a change of frame, which is implemented at the level of the partition function $\CZ$ by a generalized Fourier
transform~\cite{Aganagic:2006wq}. There are canonical choices of frame associated to special points in moduli space. In particular,
the large radius frame is adapted to the large radius point in moduli space, and in this frame
the holomorphic free energies
have an expansion in terms of Gromov--Witten invariants. There is also a conifold frame adapted to the conifold point of
the moduli space. We also note that the expression of the free energies in terms of the propagators $S^{ab}$ is independent of the
frame. However, the holomorphic limit of the propagators $S^{ab}$, which we will
denote by $\CS^{ab}$, does depend on a choice of frame, and different choices of this holomorphic
limit in the solution of the HAE give the different, frame-dependent holomorphic free energies.

In the case of local CY manifolds the holomorphic ambiguity can be fixed in many cases by using the behavior of the free energies at
special points in moduli space \cite{hkr}. Let us review this
behavior in the case of the refined topological string.
We first consider the behavior at the conifold locus. Let $t_c$ be an appropriately normalized
flat coordinate vanishing at this locus, and parametrizing a normal direction to it.
In the conifold frame, the free energy has the following behavior as $t_c \to 0$ \cite{hk,kw}
\be
\label{rf-con}
\CF_g(\bt;\mb) ={c_g(\mb) \over t_c^{2g-2}} + \CO(1), \qquad g \ge 2,
\ee
where the coefficient $c_g(\mb)$ is given by
\be
c_g(\mb)= -(2g-3)! \sum_{m=0}^g \widehat B_{2m} \widehat B_{2g-2m} \mb^{2(2m-g)}.
\ee
In this formula
\be
\widehat B_m= \big(2^{1-m}-1\big)  {B_m \over m!}
\ee
and $B_m$ is the Bernoulli number. In the unrefined limit $\mb=1$ one has
\be
\label{cgb1}
c_g(1)= {B_{2g} \over 2g (2g-2)},
\qquad
g\ge 2,
\ee
and recovers the well-known universal conifold behavior of the standard topological string~\cite{gv-conifold}.

Another universal result concerns the behavior of the refined free energies at large radius and in the large radius frame.
It is possible to generalize the Gopakumar--Vafa integrality structure~\cite{gv} to a refined version \cite{hiv,hk}.
Let us introduce
\be
\epsilon_{L,R} = {\epsilon_1 \mp \epsilon_2 \over 2} .
\ee
In terms of the characters $\chi_{j} (y)$ defined in \eqref{defcharacter},
the total free energy is then given, up to a~cubic polynomial in $\bft$, by a sum of the form
\be
\label{refBPS}
\CF({\bft}; g_s, \mb)=\sum_{w,\bfd}\sum_{j_L, j_R}
\frac{\chi_{j_L}\big( {\rm e}^{\I w \epsilon_L} \big)
\chi_{j_R} \big( {\rm e}^{\I w \epsilon_R}\big)}{4 \sin \big( {g_s w \mb \over2} \big)
\sin \big( {g_s w \over 2 \mb} \big) }
N^{j_L, j_R}_{\bfd} Q_{\bfd}^{ w} ,
\ee
where $Q_{\bfd}=\re^{-\bfd \cdot \bft}$ and $N^{j_L, j_R}_{\bfd}$ are integers (here, $j_L,j_R\in \IZ^+/2$).
The integers $N^{j_L, j_R}_{\bfd}$ count
BPS states with charge ${\bfd}$ transforming with spin $(j_L,j_R)$
under the little group ${\rm SU}(2)_L\times {\rm SU}(2)_R$ in~5 dimensions,
and are sometimes called BPS invariants.
When $\mb=1$, we recover the integrality structure of the standard topological string free
energy. In particular,
we have the following relation between the genus zero Gopakumar--Vafa invariants \smash{$n^{(0)}_{\bfd}$} and the BPS invariants
\be
\label{g0GV}
n_{\bfd}^{(0)}= \sum_{j_L, j_R} d_L d_R N^{j_L, j_R}_{\bfd},
\qquad d_{L,R}= 2 j_{L,R}+1.
\ee
The refined (or motivic) Donaldson--Thomas invariants $\Omega({\bfd},y)$
in the large volume limit are characters for the diagonal ${\rm SU}(2)$ action:
\be
\label{OmfromN}
\Omega({\bfd},y)= \sum_{j_L, j_R} \chi_{j_L}(y)  \chi_{j_R}(y)  N^{j_L, j_R}_{\bfd}.
\ee
In particular, they reduce to \eqref{g0GV} in the unrefined limit $y\to 1$.
Unlike the BPS invariants~$N^{j_L, j_R}_{\bfd}$,
which are only defined when ${\bfd}$ is a curve class, the refined DT invariants
$\Omega(\gamma,y)$
are defined for any vector $\gamma\in K_0(X)$, i.e., for arbitrary D6-D4-D2-D0 brane charge in type~IIA,
or D5-D3-D1-D(-1) charge in type IIB.
For later purposes, it will be convenient to decompose~$\Omega(\gamma,y)$ as a sum of ${\rm SU}(2)$ characters,
\be
\label{defOmj}
\Omega(\gamma,y)
 := \sum_j \chi_{j}(y)  \Omega_{[j]}(\gamma),
\ee
where the integers $\Omega_{[j]}(\gamma)$
count BPS multiplets of angular momentum $j\in\IZ/2$,
and $y$ is the corresponding fugacity parameter.
We note that the representation \eqref{defOmj} implies
that $\Omega(\gamma,y)$ is a Laurent polynomial in $y$, invariant under Poincar\'e duality $y\mapsto 1/y$.
As emphasized in \cite{Mozgovoy:2020has}, for the standard definition of DT invariants using cohomology
with compact supports, Poincar\'e duality is broken when $X$ is non-compact, ultimately due
to the non-invariance of the attractor indices counting D0-branes. Here we assume that
the DT invariants which are pertinent for the refined topological string are in fact invariant under Poincar\'e duality.

\section{Trans-series solutions and resurgent structure}
\label{sec_trans}

In order to understand the resurgent structure of the refined topological string, we
follow the strategy of \cite{cesv2,cesv1, cms, Gu:2023mgf, gm-multi}
and look for trans-series solutions to the HAE. These define non-perturbative sectors of the theory, but they require
boundary conditions, as in the perturbative case. As first pointed out in \cite{cesv2,cesv1},
we can obtain these boundary conditions
by looking at the large order behavior of the genus expansions near special points in moduli space.
It turns out that the first aspect of this procedure, namely the construction of formal trans-series solutions to the HAE,
is essentially identical to the unrefined case. Therefore, we will be rather succinct
and refer the reader to \cite{Gu:2023mgf,gm-multi}. The second aspect of the procedure, namely the analysis of the boundary
conditions, has some new ingredients, and we will analyze it in more detail.

\subsection{Trans-series solutions}

As in \cite{Gu:2023mgf,gm-multi}, we want to construct multi-instanton trans-series solutions for the free energy and partition function. Let us
denote by $Z^{\rm NP}$ a non-perturbative correction to the perturbative partition function $Z$ (i.e., involving an exponentially small
dependence on the string coupling constant). We will also introduce the ``reduced" partition function
\be
Z_{\rm r}={Z^{\rm NP} \over Z},
\ee
where $Z$ denotes the perturbative partition function. Following \cite{cesv2,cesv1, Gu:2023mgf, gm-multi}, we will
write down an equation satisfied by the reduced partition function $Z_{\rm r}$ as a consequence of the HAE. To do that, we
introduce a pair of $\CA$-dependent operators, $\mW$ and $\oD$.
In the local case these operators are defined as follows.
Let us introduce
\be
T^a = g_s \partial_b \CA \big(S^{ab}- \CS_{\CA}^{ab} \big).
\label{Ta}
\ee
Here, $ \CS_{\CA}^{ab}$ is the holomorphic limit of the propagator in a so-called $\CA$-frame, i.e., a symplectic frame in
which the action $\CA$ is a linear combination of the flat coordinates $t^a$. Then
\be
\label{DTa}
\oD = T^a {\partial \over \partial z^a} .
\ee
To define the operator $\mW$, we first introduce the derivation $\omega_S$ defined as
\be
\omega_S= {1\over g_s^2} T^a T^b {\partial \over \partial S^{ab}} .
\ee
Then $\mW$ is given by
\be
\mW= \omega_S- \sum_{g \ge 1} \big[ \oD \big( g_s^{2g-2} F_g \big) \big] \oD,
\ee
where in the refined case the free energies are $\mb$-dependent. As shown in \cite{Gu:2023mgf,gm-multi}, as a consequence
of the HAE, the reduced partition function satisfies the linear equation
\be
\label{wf1-eq-Z}
\mW Z_{\rm r}= {1 \over 2} \oD^2 Z_{\rm r}.
\ee
We will consider a special class of solutions of the form
\be
Z_{\rm r}=\exp(\Sigma_{\lambda}),
\label{esigma}
\ee
where $g_s \Sigma_{\lambda}$ is a formal power series in $g_s$. More precisely, we will write
\be
\label{Zas}
\Sigma_{\lambda} =-{\lambda \over g_s} \CA+ \CO\big(g_s^0\big),
\ee
so that \eqref{esigma} is a non-perturbative, exponentially small quantity. In this equation, $\lambda$ is an
arbitrary constant, and following \cite{cesv2,cesv1,dmp-np,Gu:2023mgf} we will assume that $\CA$ is an integer period of the mirror
CY manifold. By analogy with instanton physics, we will sometimes call $\CA$ an instanton action.
We note that, since \eqref{wf1-eq-Z} is linear, arbitrary linear combinations of the
basic solutions \eqref{esigma} are also solutions. As we will see, these solutions will be enough to construct the relevant trans-series
for the refined topological string.

Since $\oD$ is a derivation, the linear equation \eqref{wf1-eq-Z} leads to the following operator equation for~$\Sigma_\lambda$:
\be
\mW \Sigma_\lambda = \frac{1}{2}\big(\oD^2\Sigma_\lambda + (\oD\Sigma_\lambda)^2\big).
\label{eq:WSig}
\ee
In \cite{Gu:2023mgf,gm-multi}, this equation was solved as follows. Consider the formal series
\begin{equation}
\label{g-def}
G= {\CA \over g_s} + \sum_{g \ge 1} \oD \big( g_s^{2g-2} F_g \big),
\end{equation}
where $\CA$ is the action appearing in \eqref{Zas}. Let us now assume that $G$ satisfies the equation
\be
\label{G-eq}
\mW G= {1\over 2} \oD^2 G.
\ee
Then
\begin{equation}
\label{sigma-sol}
\Sigma_\lambda = \mO^{(\lambda)} G,
\ee
where
\be
\mO^{(\lambda)}= \sum_{k \ge 1} {(-\lambda)^{k} \over k!} \oD^{k-1} ,
\label{eq:Sig}
\end{equation}
satisfies \eqref{eq:WSig}.
Therefore, it suffices to check that \eqref{G-eq} is still true in the
refined case. After using the HAE, one finds that
\eqref{G-eq} holds if and only if
\be
\label{g-f1ref}
\omega_S \left (\oD F_1 \right)- \oD\left({\CA \over g_s} \right)\oD F_1 -
{1\over 2} \oD^2 \left({\CA \over g_s} \right)=0.
\ee
This equation is true in the conventional
topological string, when $F_1=F_1(\mb=1)$, as one can check by using \eqref{F1prop}. In particular,
it holds for any choice of holomorphic ambiguity \smash{$f^{(1)}_i(\bfz)$}. It is easily checked to be true
in the refined case as well: since $F_1(\mb)$ differs from $F_1(\mb=1)$ in a purely holomorphic function
of the moduli, as noted after \eqref{full-g1}, it follows immediately that~\eqref{g-f1ref} must also
be true for $F_1(\mb)$. This can also be checked by a direct calculation.

Boundary conditions for trans-series solutions to the HAE are obtained by considering their holomorphic
limit in an $\CA$-frame. In this limit, the propagator $S^{ab}$ has to be set to $\CS^{ab}_\CA$,
 and the~$T^a$ in \eqref{Ta}, as well as the operator $\oD$, vanish. It is easy to see from the formulae above
that~$\Sigma_\lambda$, when evaluated in the $\CA$-frame, is given by
\be
\label{sig-A}
\Sigma_{\lambda,  \CA}=-\lambda  {\CA \over g_s}.
\ee
The solution \eqref{sigma-sol} involves the full, non-holomorphic propagators,
but in practice one wants to understand its holomorphic limit. This goes as follows.
First, we write the action $\CA$ as a linear combination of periods\footnote{For compact CY,
there is an additional contribution $-\frac{c^0}{2\pi} (\CF_0-2 t^a\partial_a \CF_0)$, corresponding at large radius to the D5-brane
charge in IIB, or D6-brane charge in IIA, but this term
is absent when $X$ is non-compact.}
\be
\label{A-periods}
\cA=-\ri c^a {\partial \CF_0 \over \partial t^a}+ 2 \pi d_a t^a + 4 \pi^2 \ri d_0.
\ee
For example, in the large radius frame, we have that $t^a= -\log(z^a)+\cdots$ are the complexified K\"ahler parameters,
and the genus zero free energy $\CF_0$ behaves as\footnote{Here $\kappa_{abc}$ is a rational number,
which plays the role of the triple intersection number for a non-compact CY threefold. Note
that in our conventions, monodromies around the large radius point induce shifts $t^a\mapsto t^a +2\pi\I\epsilon^a$ with $\epsilon^a\in\IZ$.}
\be \CF_0= {\kappa_{abc} \over 3!} t^a t^b t^c+ \CO\big(\re^{-t^a}\big),
\ee
as $t^a\to +\infty$. In \eqref{A-periods}, $(c^a,d_a,d_0)$ are integers, and in the large radius frame they correspond
to D3-D1-D(-1) charges
in type IIB set-up, or D4-D2-D0 charges in type IIA. The action $\CA$ is then identified with
the central charge $Z(\gamma)$, up to an overall factor $4\pi^2\I$.

When not all $c^a$ vanish at once, one defines a new prepotential by
\be
\cA=-\ri c^a {\partial \widetilde \CF_0 \over \partial t^a}.
\ee
It differs from the conventional prepotential at most in a quadratic polynomial in the flat coordinates $t^a$.
Then, in the holomorphic limit, one has
\be
\oD \rightarrow -\ri g_s c^a {\partial \over \partial t^a} ,
\ee
and $\Sigma_\lambda$ becomes
\be
\Sigma_\lambda \rightarrow \tilde \CF(\bft+\ri \lambda g_s\bfc ; g_s, \mb)-\tilde \CF(\bft; g_s, \mb),
\ee
where
\be
\label{tildef}
\tilde \CF( {\bft}; g_s, \mb)= g_s^{-2} \tilde \CF_0({\bft})+ \sum_{g \ge 1} g_s^{2g-2} \CF_g( {\bft}; \mb),
\ee
i.e., it differs from the conventional free energy only in the genus zero part.

\subsection{Boundary conditions from the conifold}

Let us now discuss the boundary conditions for the trans-series. These follow from the
behavior of the free energies, in appropriate
frames, at special points in the moduli space. This behavior determines the large order
asymptotics of the free energies, and this leads
in turn to multi-instanton trans-series.

Let us first review how the conifold boundary condition for the trans-series is determined in the case of
the standard topological string with $\mb=1$ \cite{cesv2,cesv1}.
 The free energies in the conifold frame have the following behavior near the conifold locus,
\be
\label{fg-consing}
\CF_g (\bt; \mb=1) = {B_{2g} \over 2g (2g-2)} t_c^{2-2g}+ \CO(1),\qquad g\ge 2,
\ee
which follows from \eqref{rf-con}, \eqref{cgb1}.
This singular behavior also determines
the growth of the free energies at large genus
close to the conifold locus where $t_c=0$. To compute the large order behavior,
one can use the following representation of the Bernoulli numbers
\be
\label{bernoulli-as}
B_{2g} = (-1)^{g-1} {2 (2g)! \over (2 \pi)^{2g}} \sum_{\ell \ge 1}\ell^{-2g}
\ee
to obtain the formula
\be
\label{lobern}
{B_{2g} \over 2g (2g-2)} t_c^{2-2g}= {1\over 2 \pi^2} \Gamma(2g-1) \sum_{\ell \ge 1}
(\ell \CA )^{1-2g}{\CA \over \ell} \left( 1 + {1 \over 2g-2} \right), \qquad g\ge 2,
\ee
where
\be
\label{ac-gen}
\CA= 2 \pi \ri t_c.
\ee
 Let us note that, in local CY manifolds, conifold flat
coordinates are linear combinations of large radius periods $\partial F_0 /\partial t^i$
and of constant periods (see, e.g., \cite{Codesido:2015dia}), therefore \eqref{ac-gen} is the central charge of a D4-D0 BPS state.

The formula \eqref{lobern} makes manifest the asymptotic factorial growth of the left-hand side for~${g \gg 1}$, and it includes all corrections to the large $g$ behavior. Standard arguments in the
theory of resurgence relate the large order behavior of
asymptotic series to exponentially small corrections
(see, e.g., \cite[Section 3.3]{mmbook}). The basic idea is that factorial growth leads
to singularities in the Borel transform of the asymptotic series, and these in turn lead to
exponentially small discontinuities in lateral Borel resummations. By using these arguments, one finds that
the Borel transform of the perturbative series given by the left-hand side of \eqref{lobern} has singularities at the point $\ell \CA$,
where $\ell \in \IZ \backslash \{0\}$, and $\CA$ is given by \eqref{ac-gen}. In addition, one finds that
each of these singularities leads to an $\ell$-th instanton amplitude of the Pasquetti--Schiappa form \cite{ps09},
 \be
 \label{ps-ell}
 \CF_\CA^{(\ell)} ={1\over 2 \pi} \left[{1\over \ell} \left( {\CA \over g_s} \right)+ {1\over \ell^2} \right] \re^{-\ell \CA /g_s}, \qquad \ell \in \IZ \backslash \{0\}.
 \ee
This gives the $\ell$-th order trans-series which will provide boundary conditions in the conifold frame.

Another way of finding the trans-series corresponding to the large genus asymptotics of \eqref{lobern}
is to write an integral formula for the coefficient, of the form
\begin{align}
(-1)^{g-1} {B_{2g} \over 2g (2g-2)}&= {1\over 2 \pi^2} \int_0^\infty {\rd z \over z^{2g-1}}
\left\{ {\rm Li}_2 \big(\re^{-2 \pi/z}\big)- {2 \pi \over z} \log\big(1- \re^{-2 \pi/z}\big) \right\}
\nonumber\\
&=-{\ri \over \pi} \int_0^\infty {\rd z \over z^{2g-1}}\log \fadone \left( -{1\over z} \right),\qquad g\ge 2,\label{ber-intfad}
\end{align}
where $\fad(z)$ is Faddeev's quantum dilogarithm (see Appendix~\ref{sec_qdilog}).
Up to the overall factor of $1/\pi$,
the integrand in the last line of \eqref{ber-intfad} gives the sum over all the multi-instantons trans-series with positive $\ell$:
\be
\sum_{\ell \ge 1} \CF_\CA^{(\ell)} =-\ri \log \fadone \left( -{\CA \over 2 \pi g_s} \right).
\ee
This function determines the structure of the Stokes automorphism, as explained
in \cite{Iwaki:2023cek}, and in line with what was obtained in \cite{Alexandrov:2015xir}.

Let us now consider the generalization of the above result to the refined
topological string.\footnote{See \cite{Alim:2022oll,Grassi:2022zuk} for earlier studies of the resurgent structure
of the refined topological string on the resolved conifold.}
We need a formula for
the coefficient $c_g(\mb)$ appearing in \eqref{rf-con} which generalizes \eqref{lobern}. This formula is
\be
\label{cg-up}
c_g(\mb)={1 \over 2 \pi}\Gamma(2g-2) \sum_{\ell \ge 1} \left[ {(-1)^{\ell-1} \over \ell}
{1\over \sin \left( {\pi \ell \over \mb^2} \right)} \left( { 2\pi \ri \ell \over \mb} \right)^{2-2g}
+ \big( \mb \leftrightarrow \mb^{-1} \big) \right], \qquad g \ge 2,
\ee
and it can be derived by using Borel transform techniques, as follows. Let us define the more convenient set of coefficients:
\begin{gather}
\tilde c_0(\mb)=1, \qquad
\tilde c_1(\mb)={1\over 12} \big(\mb^2+ \mb^{-2}\big), \qquad
\tilde c_g(\mb)= (-1)^{g-1} {(2g)! \over (2g-3)!} c_g(\mb), \qquad g\ge 2,\!\!\!
\end{gather}
and its generating function
\be
\varphi(x)= \sum_{g \ge 0} \tilde c_g(\mb) x^{2g}.
\ee
The Borel transform of $\varphi(x)$ can be obtained in closed form, as
\be
\label{borelcg}
\widehat \varphi(\zeta)=
{\zeta^2 \over 4 \sin \big( {\zeta \mb \over 2} \big) \sin \big( {\zeta \mb^{-1} \over 2} \big)} .
\ee
When $\mb^2$ is not a rational number, this function has simple pole singularities at
\be
\label{sings}
\zeta= {2 \pi \ell \over \mb}
\qquad \mbox{and}\qquad
2 \pi \ell \mb,
\qquad
 \ell \in \IZ \backslash \{0\},
\ee
with residues
\be
\label{l-res}
{(2 \pi \ell)^2 (-1)^\ell \over\mb^3} {2 \over \sin \big( {\pi \ell \over \mb^2} \big)}
\ee
at the first set of poles. The residues at the second set are obtained by exchanging
$\mb \leftrightarrow \mb^{-1}$. The standard connection between singularities
of the Borel transform and large order asymptotics gives then the analogue of the formula \eqref{lobern} for the coefficients $\tilde c_g(\mb)$, and by going back to the original coefficients $c_g(\mb)$ we obtain \eqref{cg-up}. In our derivation, we have
assumed that $\mb^2$ is not rational. When $\mb^2 \in \mathbb{Q}$, we have singularities in the denominators in
\eqref{cg-up}. One can verify though that the singularities cancel between the two summands related
by $\mb \leftrightarrow \mb^{-1}$, as noted in \cite{ho}. The resulting expression for Faddeev's
quantum dilogarithm when $\mb^2$ is rational can be found in \cite{garkas}.

We can now use the expression \eqref{Phibsum} for Faddeev's quantum dilogarithm
to obtain the following generalization of \eqref{ber-intfad}
\be
\label{ref-id}
(-1)^{g-1} c_g(\mb)=-{\ri \over \pi} \int_0^\infty {\rd z \over z^{2g-1}}\log \fad \left( -{1\over z} \right).
\ee
Using \eqref{Phibsum}, we conclude that the relevant trans-series are given by
\begin{gather}
\label{con-ts}
\CF_{\CA, \mb}^{(\ell)}=
{(-1)^{\ell-1} \over \ell} {\re^{-{\ell \CA \over \mb g_s}} \over 2 \sin \big( {\pi \ell \over \mb^2} \big)} ,
\qquad
\CF_{\CA, \mb^{-1}}^{(\ell)}=
{(-1)^{\ell-1} \over \ell} {\re^{-{\ell \CA \mb \over g_s}} \over 2\sin \big( {\pi \ell \mb^2 } \big)} ,
\end{gather}
and
\be
\label{logfad}
\sum_{\ell \ge 1} \big( \CF_{\CA, \mb}^{(\ell)}+\CF_{\CA, \mb^{-1}}^{(\ell)} \big)
= -\ri \log \fad \left( -{\CA \over 2 \pi g_s} \right).
\ee
In particular, the trans-series involves two different actions $\cA/\mb$ and $\mb\cA$, as first
observed in~\cite{Grassi:2022zuk}.

Let us now use these boundary conditions to obtain appropriate trans-series in an arbitrary frame.
After exponentiating, we find
\be
\label{zrA}
Z_{{\rm r}, \CA}= \exp \bigg[\sum_{\ell \ge 1} \bigl( \CF_{\CA, \mb}^{(\ell)}
+\CF_{\CA, \mb^{-1}}^{(\ell)} \bigr) \bigg]= \sum_{n,m \ge 0} C_{n,m}
\exp \left( -{n \CA \over \mb g_s} - {m \mb \CA \over g_s} \right),
\ee
where $C_{n,m}$ are constants (depending on $\mb$) which can be read from
the representation \eqref{Phibsum} of $\log \fad \left( -{1\over z} \right)$.
We have, for example,
\be
C_{1,0}= {1\over 2 \sin \big( {\pi \ell \over \mb^2} \big)} ,
\qquad
C_{0,1}= {1\over 2 \sin \big( {\pi \ell \mb^2} \big)} .
\ee
We note that each of the terms in the sum of the right-hand side in \eqref{zrA} is of the form $\exp(\Sigma_\lambda)$, with
\be
\lambda= {n \over \mb}+ m \mb.
\ee
Therefore, we find by linearity
\be
Z_{\rm r}= \sum_{n,m \ge 0} C_{n,m} \exp\big( \Sigma_{ {n \over \mb}+ m \mb} \big),
\ee
with holomorphic limit
\be
\label{holz}
\CZ_{\rm r}=\sum_{n,m \ge 0} C_{n,m}
\exp\big[ \tilde \CF(\bft+ \ri g_s\bfc (n/\mb+ m \mb ) )-\tilde \CF(\bft) \big].
\ee
For example, the one-instanton amplitude is
\be
\CF^{(1)}= {1\over 2 \sin \big( {\pi \over \mb^2} \big)} \exp\big[\tilde\CF(\bft+\ri g_s\bfc /\mb)-\tilde \CF(\bft) \big]
+ \big( \mb \leftrightarrow \mb^{-1} \big).
\ee
When expanded in $g_s$, we find
\begin{align}
\CF^{(1)}={}& {\re^{-{\CA \over \mb g_s}} \over 2 \sin \big( {\pi \over \mb^2} \big)}
\exp \left( -{c^a c^b \over 2 \mb^2} \tau_{ab} \right)
\left[ 1-\ri \left( {c^a \over \mb} {\partial \CF_1 (\mb) \over \partial t^a}
- {1\over 6 \mb^3}  c^a c^b c^e C^t_{abe} \right)g_s+ \cdots \right]\nonumber
\\
& + \big( \mb \leftrightarrow \mb^{-1} \big),\label{oneinst-ser}
\end{align}
where
\be
\tau_{ab}= {\partial^2 \widetilde \CF_0 \over \partial t^a \partial t^b} ,
\qquad
C^t_{abe} = {\partial^3 \CF_0 \over \partial t^a \partial t^b \partial t^e} .
\ee
Note that $\CF_1(\mb)$ is the holomorphic limit of the full, $\mb$-dependent genus one amplitude \eqref{full-g1}. We expect that
the trans-series \eqref{oneinst-ser} will control the asymptotic behavior of $\CF_g({\bft}; \mb)$ at large genus,
not far from the conifold locus. We will test this numerically in Section \ref{sec-numerics}.

\subsection{Boundary conditions from large radius}

As we have seen in the last section, the behavior of the refined topological string
predicts the existence of a Borel singularity given (up to a normalization) by the conifold flat coordinate,
and leads to an explicit form for the corresponding multi-instanton trans-series. This can be
transformed to an arbitrary frame, leading to \eqref{holz}.
In \cite{Gu:2023mgf}, it was shown that, by using the Gopakumar--Vafa formula \cite{gv}, one
can obtain information on the Borel singularities and their trans-series near the large radius point. We will now generalize
the argument in \cite{Gu:2023mgf} to the refined case.

The starting point is the representation \eqref{refBPS} of the topological free energy
in terms of the BPS invariants. Let us introduce
\be
\mb_\pm= \mb\pm \mb^{-1},
\ee
and the coefficients $s_g^{j_L, j_R} (\mb)$ defined in terms of the generating function
\be
\label{sg-def}
\frac{\chi_{j_L}\big( {\rm e}^{\frac{\I x\mb_+}{2} }\big) \chi_{j_R} \big( {\rm e}^{\frac{\I x \mb_-}{2}}\big)}
{4 \sin \big( {x \mb\over2} \big)
\sin \big( {x \over 2\mb} \big) }= \sum_{g \ge 0} s_g^{j_L, j_R}(\mb)  x^{2g-2}.
\ee
We have, for example,
\begin{gather}
s_0^{j_L, j_R} (\mb)= d_L d_R ,
\qquad
s_1^{j_L, j_R} (\mb)={d_L d_R \over 24} \bigl[ \mb^2+ \mb^{-2}+ \big(1- d_L^2\big) \mb_+^2 + \big(1- d_R^2\big) \mb_-^2 \bigr].
\end{gather}
By expanding both sides of \eqref{refBPS} in powers of $g_s$, we obtain the refined multicovering formula
\be
\CF_g ({\bft};\mb)= \sum_{\bfd} \sum_{j_L, j_R}N^{j_L, j_R}_{\bfd} s_g^{j_L, j_R}(\mb)
 \Li_{3-2g} (Q_{\bfd}).
\ee
Now, as in \cite{Gu:2023mgf}, we use
\be
\label{polylog}
\sum_{n \in \IZ} {1\over (2 \pi n - \ri t)^{m}}=
{\ri^m \over (m-1)!} \Li_{-m+1} \big(\re^{-t}\big), \qquad m \ge 2,
\ee
to write the free energy as
\be
\label{fg-multicov}
\CF_g({\bft};\mb)= \sum_{\bfd}\sum_{n \in \IZ} \sum_{j_L, j_R} N^{j_L, j_R}_{\bfd} r_g^{j_L, j_R} (\mb)
 \CA_{ {\bfd}, n} ^{2-2g}, \qquad g \ge 2,
\ee
where
\be
r_g^{j_L,j_R} (\mb)= (-1)^{g-1} (2 \pi \ri)^{2g-2} (2g-3)!  s_g^{j_L, j_R} (\mb)
\ee
and
\be
\label{action-lr}
\CA_{ {\bfd}, n}=2 \pi {\bfd} \cdot {\bft} + 4 \pi^2 \ri n  .
\ee

We want to obtain the large genus behavior of \eqref{fg-multicov},
and extract the corresponding trans-series.\footnote{In the case of the resolved conifold,
the Borel transform of the total free energy was calculated in \cite{Grassi:2022zuk}.}
To do this, we need to know the large order behavior of the
coefficients $r_g^{j_L,j_R} (\mb)$. This can be obtained by using
an argument similar to the one employed in equations~\eqref{borelcg}--\eqref{l-res}
to derive \eqref{cg-up}. One
notes first that the series \eqref{sg-def} can be regarded as
the Borel transform of the divergent series with coefficients $(2g)!  s_g^{j_L, j_R}$. The singularities of this
Borel transform are also at \eqref{sings}, and the calculation of the residues is straightforward. Introducing the variables
\be
y_\mb = -\re^{ \pi \ri \mb^2}, \qquad \ty_{\mb}= -\re^{- \pi \ri/\mb^2},
\label{defyb}
\ee
one finds the following asymptotic expansion for $g \gg 1$
\be
\label{rg-up}
r_g^{j_L, j_R} (\mb)
\sim {1 \over 2 \pi}\Gamma(2g-2) \sum_{\ell \ge 1}
\left[ {(-1)^{\ell -1} \over \ell}
\frac{ \chi_{j_L} \big(\ty_\mb^{\ell}\big)
 \chi_{j_R} \big(\ty_\mb^{\ell}\big) }
 {\sin \big( {\pi \ell \over \mb^2} \big)}
\left( { \ell \over \mb} \right)^{-2g+2}
+ \big( \mb \rightarrow \mb^{-1} \big) \right].
\ee
We conclude that there is a sequence of Borel singularities at
\be
\label{seqsing}
{\ell \over \mb} \CA_{{\bfd},n}
\qquad \mbox{and}\qquad
{\ell \mb} \CA_{{\bfd},n},
\qquad \ell \in \IZ_{>0}, \qquad n \in \IZ,
\ee
and the corresponding trans-series are
\begin{gather}
 \CF_{\CA, \mb, j_L, j_R}^{(\ell)}= {(-1)^{\ell-1} \over \ell}
{ \chi_{j_L} \big(\ty_\mb^{\ell}\big) \chi_{j_R} \big(\ty_\mb^{\ell}\big)
\over 2\sin \big( {\pi \ell \over \mb^2} \big)}
 \re^{-{\ell \CA \over \mb g_s}},\nonumber\\
 \CF_{\CA, \mb^{-1}, j_L, j_R}^{(\ell)}= {(-1)^{\ell -1} \over \ell}
{ \chi_{j_L} \big(y_\mb^{\ell}\big) \chi_{j_R} \big(y_\mb^{\ell}\big)
\over 2 \sin \big( {\pi \ell \mb^2} \big)}
 \re^{-{\ell \mb\CA \over g_s} },
\end{gather}
where $\CA= \CA_{ {\bfd}, n}$. When $j_L=j_R=0$,
we recover \eqref{con-ts}. The sum over all multi-instanton sectors gives
\be
\label{logfadj}
\sum_{\ell \ge 1} \big( \CF_{\CA, \mb, j_L, j_R}^{(\ell)}+
 \CF_{\CA, \mb^{-1}, j_L, j_R}^{(\ell)} \big)= -\ri \log \fad^{[j_L, j_R]}
 \left( -{\CA \over 2 \pi g_s} \right),
\ee
where the series on the right-hand side is the following generalization
of Faddeev's quantum dilogarithm
\begin{gather}
\log \fad^{[j_L, j_R]} \left( -{\CA \over 2 \pi g_s} \right) \nonumber\\
\qquad=
{1\over \ri} \sum_{\ell \ge 1} {(-1)^{\ell} \over \ell} \left[
{\chi_{j_L} \big(\ty_\mb^{\ell}\big)
\chi_{j_R} \big(\ty_\mb^{\ell}\big) \over 2\sin \big( {\pi \ell \over \mb^2} \big)}
 \re^{-{\ell \CA \over \mb g_s}}
+{\chi_{j_L} \big( y_\mb^{\ell}\big)
\chi_{j_R} \big( y_\mb^{\ell}\big) \over 2\sin \big( {\pi \ell \mb^2 } \big)}
 \re^{-{\ell \CA \mb \over g_s}}\right].\label{gen-fad}
\end{gather}
When $j_L=j_R=0$, we recover the conventional
quantum dilogarithm. In fact, \eqref{gen-fad} can be expressed in terms of a more elementary function \smash{$\fad^{[j]} (z)$}
defined in \eqref{defPhibj}.
Indeed, from the representation of \smash{$\fad^{[j]} (z)$} given in the first line of \eqref{reprPhibj},
 one finds the following identity
\be
\log \fad^{[j_L, j_R]} (z)=
\sum_{j=|j_L-j_R|}^{j_L+j_R} \log \fad^{[j]} (z),
\ee
where the sum runs over half-integer $j$ such that $j-j_L-j_R$ is integer.
The third line in \eqref{reprPhibj} also shows that
\smash{$\fad^{[j]}(z)$} can be expressed as a product of the ordinary quantum dilogarithms with shifted arguments.

The full trans-series corresponding to all the singularities near large radius is
a sum of trans-series of the form \eqref{gen-fad},
\begin{gather}
 -\ri \sum_{\bfd} \sum_{n \in \IZ} \sum_{j_L, j_R} N_{\bfd}^{j_L, j_R}
\log \fad^{[j_L, j_R]} \left( -{\CA_{{\bfd}, n} \over 2 \pi g_s} \right)\nonumber\\
\qquad= -\ri \sum_{\bfd} \sum_{n \in \IZ} \sum_{j} \Omega_{[j]}(\bfd)
\log \fad^{[j]} \left( -{\CA_{{\bfd}, n} \over 2 \pi g_s} \right)\label{trans-LR}
\end{gather}
where
\be
\Omega_{[j]}(\bfd) = \sum_{j_L, j_R \atop |j_L-j_R|\leq j \leq j_L+j_R} N_{\bfd}^{j_L, j_R} \ .
\ee
This supports the identification
of the Stokes constant near the large radius point with the refined DT invariants $\Omega_{[j]}(\bfd)$
advocated below \eqref{defOmj}. Note that
the Stokes constants are common to all singularities \eqref{seqsing}
with different values of $\ell$ and $n$.
In the limit $\mb \rightarrow 1$, upon using \eqref{g0GV}, we recover the result of \cite{Gu:2023mgf}
identifying the Stokes constants near large radius with the genus 0 GV invariants,
\be
-\ri \sum_{\bfd} \sum_{n \in \IZ} \Bigg( \sum_{j_L, j_R} d_L d_R N_{\bfd}^{j_L, j_R} \Bigg) \log \fadone
\left( -{\CA_{{\bfd}, n} \over 2 \pi g_s} \right)
= -\ri \sum_{\bfd} \sum_{n \in \IZ} n_{\bfd}^{(0)} \log \fadone \left( -{\CA_{{\bfd}, n} \over 2 \pi g_s} \right).
\ee

\subsection{Stokes automorphism in the refined case}

From the above results, and using the methods of \cite{Iwaki:2023cek}, one can easily obtain the
Stokes automorphism corresponding to the trans-series \eqref{logfad}, \eqref{logfadj}. We recall that, in the theory
of resurgence, the Stokes automorphism can be obtained from the discontinuity of Borel resummations
as we cross a ray of singularities. This discontinuity is expressed as a formal linear combination of
resummed trans-series associated to the singularities, whose coefficients are the Stokes constants. The Stokes automorphism
is given by this formal linear combination of trans-series. We refer, e.g., to \cite{Iwaki:2023cek}
for additional background and references on the Stokes automorphism and related aspects of the theory of resurgence.

We found that, for the topological free energies in the large radius frame,
the leading Borel singularities near the large radius point
are at $\ell \mb^{\pm 1} \CA_{{\bfd}, n}$, with $\ell \in \IZ \backslash{0}$, corresponding to D2-D0 branes in type IIA, or D1-D(-1) instantons in IIB.
We note that, for general $\mb$ and $\ell >0$, we have two rays of singularities. For convenience, we consider
the Stokes automorphism associated to the discontinuity as we cross both rays.\footnote{One
could in principle consider separately the Stokes automorphism for the ray containing the singularities~${\ell \mb \CA_{{\bfd}, n}}$,~${\ell >0}$, and the Stokes automorphism
for the ray containing $\ell \mb^{-1} \CA_{{\bfd}, n}$, $\ell >0$.
However, when $\mb$ is rational, they are both singular separately, and it is only by
adding both that one obtains a finite answer.}
In this case this automorphism is purely multiplicative, and is
obtained by exponentiating the sum of multi-instantons \eqref{logfadj}
(with an additional factor of $-\ri$ due to conventions in the definition of the Stokes
automorphism). We then obtain
\be
\label{mfrakmultiLR}
\mathfrak{S}(\CZ)=\prod_{j_L, j_R} \left[\fad^{[j_L, j_R]}
\left( -{\CA \over 2 \pi g_s} \right) \right]^{-N^{j_L, j_R}_{\bfd}} \!\!\! \CZ
=\prod_{j} \left[\fad^{[j]} \left( -{\CA \over 2 \pi g_s} \right) \right]^{-\Omega_{[j]}({\bfd})} \CZ.
\ee
The case \eqref{logfad} was obtained by analyzing the behavior of the topological string
free energies in the conifold frame,
near the conifold
point where a D4-brane becomes massless (or a D3-instanton becomes of vanishing action).
The resulting Stokes automorphism is again purely multiplicative, and is in fact a particular
case of \eqref{mfrakmultiLR} in which only $j=0$ contributes with~${\Omega_{[j]}=1}$. In general,
when we consider the partition function in an $\CA$-frame, we expect the Stokes automorphism to be given by
the general expression \eqref{mfrakmultiLR}, extending what is found in the unrefined case \cite{Gu:2023mgf,gm-multi}.

Let us now consider what happens at an arbitrary frame, i.e., not necessarily an $\CA$-frame. In that case, not all $c^a$
appearing in \eqref{A-periods} vanish. As shown in
\cite{Iwaki:2023cek}, the Stokes automorphism can be obtained from the multiplicative one, after promoting the exponential
of the action $\CA$ to a~shift operator. One then finds
\be
\label{mfrakmulti2}
\mathfrak{S}(\tilde \CZ)=\prod_j \left[ \fad^{[j]}
\left( \ri g_s c^a {\partial \over \partial t^a} \right) \right]^{-\Omega_{[j]}(\gamma)} \tilde \CZ,
\ee
where we have denoted $\tilde \CZ= \re^{\tilde \CF}$ the partition function associated to the modified free energy~\eqref{tildef}.
When $\mb=1$, \eqref{mfrakmulti2} reduces to
the result of \cite{Iwaki:2023cek} for the standard topological string.
Furthermore, when $\mb^2=2$, using \eqref{bsq2}, we recover the Stokes automorphism
which appears in the real topological string at the conifold point \cite{ms}. The relevance
of this special value for topological strings on orientifolds was anticipated in \cite{kw}.

We expect \eqref{mfrakmultiLR}, \eqref{mfrakmulti2} to give the Stokes automorphism
of the refined topological string due to arbitrary singularities at $\ell \mb^{\pm 1} \CA$, $\ell \in \IZ_{>0}$.
The Stokes constants $\Omega_{[j]}$ appearing in these formulae should be identified with the
coefficients of the character expansion of the refined DT invariant \eqref{defOmj}
associated to the corresponding BPS state, as conjectured around \eqref{SZbj}.
The formulae \eqref{mfrakmultiLR}, \eqref{mfrakmulti2} were in fact already proposed by \cite{Alexandrov:2015xir},
in a different but related context, and we will rederive them in Section \ref{refined-sec}.

\subsection[Numerical checks for X=K\_IP\^{}2]{Numerical checks for $\boldsymbol{X=K_{\IP^2}}$}
\label{sec-numerics}

As is well-known, the trans-series obtained in resurgent analysis, if correct, should control the
large order behavior of the original perturbative series. In particular, the one-instanton series~\eqref{oneinst-ser}
should give the leading asymptotic behavior in the region of moduli space
in which the leading singularity in the Borel plane is given by either $\mb \CA$ or $\CA/\mb$ (i.e., the one
which is closer to the origin). We will now verify that \eqref{oneinst-ser} indeed gives the correct answer
in the simplest non-trivial local CY,
namely, $X=K_{\IP^2}$, the total space of the canonical bundle over $\IP^2$, also known
as local $\IP^2$.
We follow the notations of \cite{gm-multi} and
denote by $z$ the standard algebraic coordinate on K\"ahler moduli space,
such that $z=0$, $z=-\frac{1}{27}$ and $z=\infty$ correspond to the large volume, conifold
and orbifold points, respectively.
For simplicity, we shall restrict ourselves to
negative values of $z$ in the large
radius region of the moduli space,
\be
\label{reg-test}
-{1\over 27}<z<0,
\ee
since with our conventions the free energies in the large radius frame are real in this interval.

\begin{figure}
\centering
\includegraphics[height=4cm]{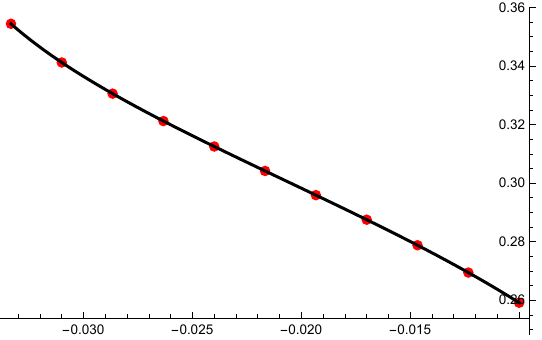} \qquad \qquad \includegraphics[height=4cm]{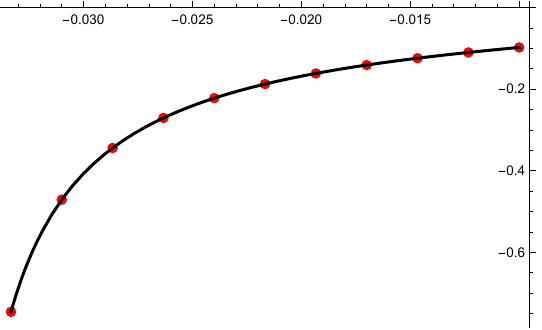}
\caption{Values of the coefficients
$\mu_0$ (left) and $\mu_1$ (right) in \eqref{Fmu0mu1}, for different values of $z$
in the range $-\frac1{27}<z<0$ and for a fixed value $\mb= \pi$, extracted from the asymptotic behaviour
of $\CF_g(\mb)$ up to~${g=35}$ (after using three Richardson transforms
to improve convergence). The numerical values are very close to the
theoretical predictions shown in black lines, with $10^{-10}$ and $10^{-9}$ accuracy
for $\mu_0$ and~$\mu_1$, respectively.}
\label{mu01plots-fig}
\end{figure}

From previous studies \cite{cesv2,cms,gm-multi}, it is known that, in most of the range in
\eqref{reg-test}, the leading Borel singularity
corresponds to the central charge of the D4-brane which becomes massless at the
conifold point $z=-1/27$, and
is given by the conifold result \eqref{ac-gen}. In particular,
the Borel singularity $\mb ^{\pm 1} \CA_{1,0}$ in \eqref{action-lr} only becomes relevant very close to the large
radius limit, when~${-10^{-6}<z<0}$. We will denote
\be
\CA_c = 2 \pi \ri t_c,
\ee
where $t_c$ is the appropriately normalized conifold
flat coordinate.\footnote{ We follow the conventions of \cite{ms} for the
normalization of $t_c$, which differs from the one used in \cite{gm-multi} by a~factor of ${\sqrt{3}}$.} Therefore, in this region the large genus behavior of the
free energies $\CF_g(\mb)$ in the large radius frame is determined
by \eqref{oneinst-ser}. Let us note that, in our conventions,
the free energy $F^{(1,0)}$ is given by
\be
\CF^{(1,0)}=-{1\over 24} \log\left( -{\Delta \over z} \right), \qquad \Delta= 1+ 27 z.
\ee
Since $b_2(X)=1$, there is only one coefficient $c^a$ in \eqref{oneinst-ser},
which is equal to $c=-3$ in the present conventions \cite{gm-multi}.
Writing $\CF^{(1)}$ as
\be
\CF^{(1)}= \re^{-{\widetilde \CA /g_s}} \left(\mu_0 + \mu_1 g_s +\cdots \right),
\ee
standard resurgent analysis predicts
\be
\label{Fmu0mu1}
\CF_g(\mb) \sim {1\over \pi} \widetilde \CA ^{-2g+2} \Gamma(2g-2)
\left( \mu_0 + {\mu_1 \widetilde \CA \over 2g-3}+ \cdots \right), \qquad g \gg 1.
\ee
We can extract numerically the value of the action $\widetilde \CA$
and of the coefficients $\mu_{0,1}$ from the perturbative free energies at sufficiently large order,
for different values of the modulus $z$ and the parameter $\mb$, and
compare those to the theoretical prediction in \eqref{oneinst-ser}. This prediction
implies in particular that, if $|\mb|>1$, we will have
\be
\label{a-b}
\widetilde \CA= {\CA_c \over \mb} .
\ee
If $|\mb|<1$, the action $\widetilde \CA$ is obtained by exchanging $\mb \leftrightarrow \mb^{-1}$ in \eqref{a-b}
(note that the sequence~$\CF_g$ is invariant under this exchange).
Computing the free energies up to $g=35$ for $\mb= \pi$ (a~convenient irrational number)
and several values of $z$, we already find excellent agreement, see Figure~\ref{mu01plots-fig}.
A~similar agreement is obtained for other values of $\mb$, including complex ones.

\section{Refined dual partition functions}
\label{refined-sec}

In this section, we propose an extension of the notion of dual partition function,
realized as certain generalized theta series encoding five-brane instantons, that incorporates the refinement.
The construction relies on the use of a non-commutative Moyal star product and its extension to the realm of contact geometry.
Our main result is that the same Stokes automorphism~\eqref{mfrakmultiLR} that governs the refined topological string
also arises as the transformation of the kernel of the refined dual partition function
induced by a non-commutative wall-crossing transformation.
Furthermore, by considering higher five-brane charge $k>1$, we obtain a vector-valued generalization of \eqref{mfrakmultiLR},
which realizes the double quantization proposed in \cite[Section~4]{Alexandrov:2015xir}.

\subsection{Five-brane instanton corrections and dual partition functions}

Let us first revisit the construction of dual partition functions and their behavior under wall-crossing
from \cite{Alexandrov:2015xir}.
This construction can be motivated by considering five-brane instanton corrections to the vector multiplet moduli
space $\CM$ arising by compactifying type IIA strings on~${X \times S_1}$ down to 3 dimensions.\footnote{Equivalently,
one can consider the hypermultiplet moduli space of type IIB string theory compactified on $X$,
or type IIA on the mirror threefold $\widehat X$.}
When $X$ is a compact CY threefold, $\CM$ is a quaternion-K\"ahler (QK) manifold of real dimension
$4b_2(X)+4$, of the form
\be
\label{CMform}
\CM \simeq \IR^+_R \times \CS\CK_{z^a} \times \CT _{\zeta^\Lambda,\tzeta_\Lambda} \times S^1_{\sigma},
\ee
where $ \IR^+$ parametrizes the radius of the circle, $\CS\CK$ the complexified K\"ahler structure on~$X$,~$\CT$ the holonomies of the Ramond gauge fields around $H_{\rm even}(X,\IZ)$, and $S^1$
the scalar Poincar\'e-dual to the Kaluza--Klein gauge field in 3 dimensions. In \eqref{CMform},
we have indicated in subscript the coordinates used to parametrize the various factors, with $a$ running over
$1,\dots,b_2:=b_2(X)=b_4(X)$, $\Lambda=0,\dots, b_2$, with $\zeta^\Lambda$ and
$\tzeta_\Lambda$ associated to $H_0\oplus H_2$ and $H_4\oplus H_6$, respectively.
Topologically, the
level sets of $R$ are principal bundles over $\CS\CK$, whose fibers are twisted tori~\smash{$\widetilde{\CT}=\IR^{2b_2+3}_{\zeta^\Lambda,\tzeta_\Lambda,\sigma}/\Heis\simeq \CT\times S^1$} where $\Heis$
is the non-commutative Heisenberg group of large gauge transformations parametrized by $\Theta=\big(\eta^\Lambda,\teta_\Lambda,\kappa\big)\in\IZ^{2b_2+3}$
and acting by
\be
\scT_{\Theta}\colon\ \big(\zeta^\Lambda,\tzeta_\Lambda,\sigma\big)\ \mapsto\
\big(\zeta^\Lambda+\eta^\Lambda, \tzeta_\Lambda+\teta_\Lambda, \sigma+2\kappa-\teta_\Lambda\zeta^\Lambda
+\eta^\Lambda\tzeta_\Lambda-\eta^\Lambda\teta_\Lambda\big) .
\label{Heisen}
\ee
The coordinate $\sigma$ can be viewed as parametrizing the phase of a section of the theta line bundle~$\mathcal{L}$
with connection $\mathcal{A}$ over the torus parametrized by $\big(\zeta^\Lambda,\tzeta_\Lambda\big)$, with curvature ${\rm d}\mathcal{A}={\rm d}\tzeta_\Lambda\wedge {\rm d}\zeta^\Lambda$~\cite{Alexandrov:2010np}.
When $X$ is non-compact,
three-dimensional gravity decouples and $\CM$ becomes a family of hyperk\"ahler manifolds
of dimension $4b_2$ parametrized by the (non-dynamical) radius~$R$,
of the form $\CS\CK_{z^a} \times \CT _{\zeta^a,\tzeta_a}$, obtained as a rigid limit of~\eqref{CMform}.

As $R\to\infty$, the QK metric on
$\CM$ is simply obtained from the special K\"ahler metric on $\CS\CK$ by the $c$-map
construction \cite{Ferrara:1989ik}, but for finite $R$ there are \smash{$\CO\big({\rm e}^{-R}\big)$} corrections
from Euclidean D-branes wrapped on even cycles in $X$ times $S^1$, and \smash{$\CO\big({\rm e}^{-R^2}\big)$}
corrections from Euclidean five-branes.\footnote{In the present context, these are Kaluza--Klein five-branes
of the form ${\rm TN}_k \times X$, where the first factor is a Taub-NUT gravitational instanton of
charge $k$ which asymptotes to $\IR^3\times S_1(R)$. Under T-duality along the circle, this becomes
a Neveu--Schwarz five-brane instanton correcting the hypermultiplet moduli space, which is the
equivalent set-up used in \cite{APP,Alexandrov:2015xir}.}
Both types of corrections must preserve the QK property of the metric, which is equivalent \cite{MR1327157}
to the existence of a complex contact structure on the twistor space ${\CZ\simeq \IP^1_{\bfz} \times \CM}$,
where the first factor parametrizes
the sphere-worth of almost complex structures on $\CM$. Such a~structure is guaranteed by the
existence of coordinate patches parametrized by complex Darboux coordinates
$\xi^\Lambda$, $\txi_\Lambda$, $\talp$ such that the contact one-form
\be
\label{contactone}
\CX=-\frac12\(\rd\talp-\xi^\Lambda \rd\txi_\Lambda+\txi_\Lambda\rd\xi^\Lambda\)
\ee
is globally well defined up to rescaling by a non-vanishing holomorphic function.
The Darboux coordinates are functions of $t$, $R$, $z^a$, $\zeta^\Lambda$, $\tzeta_\Lambda$, $\sigma$,
holomorphic in $\bfz$ in the respective patches, and can be chosen such that the Heisenberg group
acts in the same way as in \eqref{Heisen},
\be
\scT_\Theta\colon\ \big(\xi^\Lambda,\txi_\Lambda,\talp\big)\ \mapsto\
\big(\xi^\Lambda+\eta^\Lambda, \txi_\Lambda+\teta_\Lambda, \talp+2\kappa-\teta_\Lambda\xi^\Lambda
+\eta^\Lambda\txi_\Lambda-\eta^\Lambda\teta_\Lambda\big).
\label{HeisenZ}
\ee
This is an example of contact transformation, i.e., preserving the contact one-form \eqref{contactone}.
As a~result, the twistor space $\CZ$ can be obtained by gluing together algebraic twisted tori
\begin{gather*}
\smash{\widetilde{\CT}_{\IC}}=\smash{(\IC^\times)^{2b_2+3}_{\xi^\Lambda,\txi_\Lambda,\talp}/\Heis}.
\end{gather*}
By omitting the coordinate $\talp$, the latter project
to algebraic tori \smash{${\CT}_{\IC}=(\IC^\times)^{2b_2+2}_{\xi^\Lambda,\txi_\Lambda}$}, with $\IC^\times$ fiber parametrized by ${\rm e}^{-\pi\I\talp}$.

At large but finite radius $R$, D-brane instantons generate corrections of order ${\rm e}^{-2\pi R |Z(\gamma)|}$ to the QK metric on $\CM$, where $\gamma=\big(p^\Lambda,q_\Lambda\big)\in H_{\rm even}(X,\IZ)$ is the instanton charge and $Z(\gamma)$ the corresponding central charge. We denote by
$\CM_D$ the QK metric on $\CM$ incorporating all D-instanton corrections. The twistor space
$\CZ_D$
associated to $\CM_D$ can be constructed by postulating
discontinuities of the Darboux coordinates across the so-called BPS rays $\ell_\gamma$ on $\IP^1_\bfz$
(the latter being the loci where the phase of $\bfz$ coincides with the central charge $Z(\gamma)$).
Namely, one requires that across $\ell_\gamma$ they change as \cite{Alexandrov:2009zh,Alexandrov:2011ac,Alexandrov:2008gh}
\be
\scV_\gamma\colon \ \begin{aligned}
&\xi^\Lambda  \mapsto
\xi'^\Lambda = \xi^\Lambda+\frac{p^\Lambda }{2\pi\I}  \Omega(\gamma) \log(1-\cX_{\gamma}),
\\
&\txi_\Lambda  \mapsto
\txi'_\Lambda = \txi_\Lambda+\frac{ q_\Lambda }{2\pi\I} \Omega(\gamma) \log(1-\cX_{\gamma}),
\\
&\talp  \mapsto
\talp'=\talp+\frac{1}{2\pi^2} \Omega(\gamma) L_{\sigma(\gamma)}(\cX_\gamma) .
\end{aligned}
\label{VKStrans}
\ee
Here $\Omega(\gamma)$ is the generalized DT invariant counting BPS states
of charge $\gamma$ in four dimensions,
\begin{gather*}
\cX_\gamma=\sigma(\gamma){\rm e}^{-2\pi\I (q_\Lambda \xi^\Lambda-p^\Lambda\txi_\Lambda)}
\end{gather*}
are the twisted Fourier modes, with $\sigma(\gamma)$ a quadratic refinement of the symplectic intersection pairing
$\langle\gamma,\gamma'\rangle$ on the lattice of charges such that
\be
\label{twitorus}
\cX_\gamma  \cX_{\gamma'} = (-1)^{\langle\gamma,\gamma'\rangle} \cX_{\gamma+\gamma'},
\ee
and $L_\sigma(z)$ is the twisted Rogers dilogarithm
\be
L_\epsilon(z)=\Li_2(z)+\frac12 \log\big(\epsilon^{-1}z\big)\log(1-z) .
\label{Rdilog}
\ee

Forgetting the action on $\talp$, the transformation \eqref{VKStrans}
can be written in terms of the Fourier modes $\cX_\gamma$ as
\be
\scU_\gamma\colon\ \cX_{\gamma'}\ \mapsto\ \cX'_{\gamma'}=\cX_{\gamma'}(1-\cX_\gamma)^{\langle\gamma,\gamma'\rangle \Omega(\gamma)}.
\label{KStr}
\ee
This is recognized as the Kontsevich--Soibelman (KS) symplectomorphism
encoding wall-crossing transformations \cite{Gaiotto:2008cd,KS:stability},
or the Delabaere--Dillinger--Pham (DDP) formula controlling the Stokes automorphisms
for quantum periods \cite{DDP93}.
The action on the additional coordinate $\talp$ lifts $\scU_\gamma$ to a contact transformation $\scV_\gamma$.
The gluing conditions for $\talp$ further imply that ${\rm e}^{-\pi\I\talp}$ is a section of
the theta line bundle $\CL$
over the algebraic tori $\CT_\IC$ \cite{APP,Neitzke:2011za}, which descends to a
hyperholomorphic line bundle over the hyperk\"ahler space obtained from the D-instanton corrected QK space $\CM_D$
by the QK/HK correspondence \cite{Alexandrov:2011ac}.

By contrast, Neveu--Schwarz five-brane instantons induce corrections of order ${\rm e}^{-4\pi k R^2}$ and are poorly understood, beyond linear order
around the D-instanton corrected twistor space~$\CZ_D$~\cite{APP}
(see also \cite{Alexandrov:2014rca,Alexandrov:2014mfa,Alexandrov:2023hiv}). At linear
order, instanton corrections from charge $k$ five-branes are described by sections of $\CL^k$,
the $k$-th power of the theta line bundle $\CL$. In practice, this means that on local coordinate patches
they are described by functions $H_k\big(\xi,\txi,\talp\big):= {\rm e}^{-\I\pi k\talp}H_k\big(\xi,\txi\big)$ which stay invariant under
the contact transformations $\scT_\Theta$ and are mapped to each other under $\scV_\gamma$.
The functions $H_k\big(\xi,\txi,\talp\big)$ give rise to additional discontinuities of the Darboux coordinates
obtained by applying to them the operator \smash{${\rm e}^{\{H_k, \cdot \}_{1,0}}$} where $\{\cdot , \cdot\}_{1,0}$
denotes the so called contact bracket \cite{Alexandrov:2014rca,Alexandrov:2014mfa,Alexandrov:2008gh},
in the same way as the discontinuities \eqref{VKStrans} are obtained by applying ${\rm e}^{\{h_\gamma, \cdot \}_{1,0}}$
with \smash{$h_\gamma=\frac{\Omega(\gamma)}{4\pi^2} \Li_2(\cX_\gamma)$}.
The instanton corrected metric $\CM_D$ can then be derived following the procedure explained in detail
in \cite{Alexandrov:2010qdt}.

The invariance under the Heisenberg action \eqref{HeisenZ} requires that
\be
H_k\big(\xi+\eta,\txi+\teta\big)=(-1)^{k\eta^\Lambda\teta_\Lambda}
{\rm e}^{\pi\I k(\eta^\Lambda\txi_\Lambda-\teta_\Lambda\xi^\Lambda)} H_k\big(\xi,\txi\big).
\label{thetaline}
\ee
This implies the following `non-Abelian' Fourier expansion
\be
H_{k}\big(\xi,\txi,\talp\big) = \sum_{\ell^\Lambda\in \frac{\IZ^{b_2}}{|k|\IZ^{b_2}}}
\sum_{n^\Lambda \in \IZ^{b_2} + \frac{\ell^\Lambda}{k} }
{\rm e}^{-\pi\I k (\talp+\xi^\Lambda\txi_\Lambda)+2\pi\I k n^\Lambda \txi_\Lambda}
\CH_{k,\ell^\Lambda}\big(\xi^\Lambda - n^\Lambda\big) ,
\label{symseries}
\ee
where $\CH_{k,\ell^\Lambda}\big(\xi^\Lambda\big)$ is referred to as the wave-function.
The reason for this terminology is that it transforms in the metaplectic representation under a change of symplectic basis.
For example, upon exchanging the `position' coordinates $\xi^\Lambda$ with the conjugate `momenta' $\txi_\Lambda$,
one gets
\be
H_{k}\big(\xi,\txi,\talp\big) = \sum_{l_\Lambda\in \frac{\IZ}{|k|\IZ}}
\sum_{m_\Lambda \in \IZ + \frac{\ell^\Lambda}{k} }
{\rm e}^{-\pi\I k (\talp-\xi^\Lambda\txi_\Lambda)-2\pi\I k m_\Lambda \xi^\Lambda}
\CH'_{k,l_\Lambda}\big(\txi_\Lambda - m_\Lambda\big) ,
\label{symseries2}
\ee
where the wave-functions are related by Fourier transform
\be
\CH_{k,\ell^\Lambda}(\xi)
=
\sum_{m_\Lambda\in \frac{\IZ}{|k|\IZ}} {\rm e}^{-2\pi\I m_\Lambda \ell^\Lambda/k}
\int \rd\txi_\Lambda {\rm e}^{2\pi\I k \xi^\Lambda\txi_\Lambda} \CH'_{k,m_\Lambda}\big(\txi\big) .
\label{Four-gen}
\ee

For $k=1$, it was argued in \cite{APP} using S-duality in type IIB string theory
that the wave function~$\CH_{1}(\xi)$ should be identified with the topological string
partition function in real polarization~$\CZ_{\IR}(\xi)$, analytically continued to the complex domain
and evaluated on the Darboux coordinates~$\xi^\Lambda$. As a result, \eqref{symseries} becomes
proportional to the `dual partition function':
\be
H_1\big(\xi,\txi,\talp\big) = {\rm e}^{2\pi\I \alpha}
\sum_{n^\Lambda \in \IZ }
{\rm e}^{2\pi\I n^\Lambda \txi_\Lambda}\CZ_{\IR}\big(\xi^\Lambda - n^\Lambda\big) ,
\label{HonePsi}
\ee
where we introduced
\be
\alpha=-\frac12 \big(\talp+\xi^\Lambda\txi_\Lambda\big)
\label{defalpha}
\ee
such that
\be
\CX=\rd\alpha+\xi^\Lambda\rd\txi_\Lambda.
\label{cXalp}
\ee
In fact, $\CZ_{\IR}(\xi)$ can be shown to be proportional to the usual holomorphic topological string partition
function $\CZ(\bft;g_s)$ provided one identifies\footnote{The easiest way to establish these relations is to compare
the Fourier modes $\cX_\gamma$ with ${\rm e}^{-\CA/g_s}$, where the instanton action $\cA$ is given in \eqref{A-periods},
by identifying the integer valued charges $q_\Lambda$ and $(d_a,d_0)$.
Another consistency check is that in \eqref{deftau} the argument of $\CZ$ becomes $t^a-\I g_s n^a=2\pi\I(\xi^a-n^a)/\xi^0$,
which is consistent with \eqref{HonePsi} if one sets $n^0=0$. Since $n^0$ plays the role of D6-brane charge,
this restriction is indeed necessary in the non-compact case.
Finally, one can also check that the relations \eqref{ident-arg} ensure that in \eqref{Stau} the shift of $t^a$
in the argument of $\tau$ is consistent with the KS transformation of $\xi^a$ given in \eqref{VKStrans}.
}
\be
t^a=2\pi\I \frac{\xi^a}{\xi^0} ,
\qquad
g_s=\frac{2\pi}{\xi^0} .
\label{ident-arg}
\ee
Thus, the five-brane instantons turn out to be described by the unrefined version of the partition function
studied in the previous sections.

One of the main results of \cite{Alexandrov:2015xir} was to determine the
transformation property of the wave functions $\CH_{k,\ell^\Lambda}$ under the contact transformations \eqref{VKStrans}.
Taking into account the transformation of $\talp$, it is immediate to see that
\be
\label{VKStransH}
\scV_\gamma\colon\ H_k\big(\xi,\txi\big) \mapsto\ H'_k\big(\xi,\txi\big)
= {\rm e}^{\frac{k}{2\pi \I } \Omega(\gamma)  L_{\sigma(\gamma)}(\cX_\gamma)} H_k\big(\xi', \txi'\big)  .
\ee
In the pure electric case $\gamma=\big(0,q^\Lambda\big)$, it was shown that this implies
\be
\scV_\gamma\colon\ \CH_{k,\ell^\Lambda}(\xi)\ \mapsto\
\big[ \Ab^{(k)}\bigl(-q_\Lambda\xi^\Lambda,-q_\Lambda \ell^\Lambda\bigr) \big]^{-\Omega(\gamma)}
 \CH_{k,\ell^\Lambda}(\xi),
\label{trcHk}
\ee
where\footnote{The function defined in \eqref{Abkl} is the inverse of $\Ab^{(k)}(x,\ell)$ in \cite{Alexandrov:2015xir}.}
\be
\Ab^{(k)}(x,\ell)=\big(1-{\rm e}^{2\pi\I (x+\ell/k)}\big)^{\ell}
\bigl[ \Phi_1\left(\I (x+\ell/k) \right)\bigr]^{k}.
\label{Abkl}
\ee
The transformation property across generic BPS rays can be obtained by conjugating
\eqref{trcHk} by the metaplectic representation, or by extending by a rank 2 hyperbolic lattice,
see \cite[equation~(2.11)]{Alexandrov:2015xir}. For $k=1$, \eqref{trcHk} and \eqref{ident-arg} imply that the topological
string amplitude $\CZ(\bft;g_s)$ gets multiplied by $\Phi_1(- (q_a t^a+2\pi\I q_0)/g_s)$, which coincides
with the Stokes automorphism
 for the topological string partition function, as noticed in \cite{Iwaki:2023cek}.

\subsection{Refined contact structure}

Our goal is to incorporate the refinement parameter into the above construction.
In fact, a~refined version of the function $\Ab^{(k)}(x,\ell)$ was already put forward in
\cite[Section~4]{Alexandrov:2015xir}. However, it was not derived nor justified by any invariance or transformation property.
To fill this gap, we have to find a proper generalization of the dual partition function \eqref{HonePsi}
(or more generally~\eqref{symseries}) and compute the action of a refined version of
the transformation $\scV_\gamma$ \eqref{VKStrans} on its kernel.
However, as will be discussed shortly, while the refinement of the symplectomorphism $\scU_\gamma$ is well understood,
this is not so for the contact transformation $\scV_\gamma$.
Therefore, the first step is to find how to lift refined symplectomorphisms to refined contact transformations.
Since the defining property of the latter in the absence of refinement was that they preserve the contact one-form,
this can be seen as constructing a refined version of the contact structure.
This is the problem that we address in this subsection.

Physically, the standard way to introduce a refinement is to switch on an $\Omega$-background.
As was observed in \cite{Cecotti:2014wea,Gaiotto:2010be} in the gauge theory context,
its effect is to deform the Riemann--Hilbert problem defining the instanton-corrected metric on
the Coulomb branch into a
non-commutative one \cite{Barbieri:2019yya}.
This is achieved by replacing the KS symplectomorphism \eqref{KStr} by its quantum version
\begin{gather}
\hat\scU_\gamma\colon\ \cX_{\gamma'}\ \mapsto\ \cX'_{\gamma'}=\CU_\gamma \star \cX_{\gamma'} \star \CU_\gamma^{-1}.
\label{qKStr}
\end{gather}
Here $\star$ denotes the non-commutative Moyal product
\be
f \star g = f \exp\left[ \frac{\I \eps}{2\pi}\sum_{\Lambda} \left(
\overleftarrow{\p}_{\!\xi^\Lambda}\overrightarrow{\p}_{\!\txi_\Lambda}
- \overleftarrow{\p}_{\!\txi_\Lambda}\overrightarrow{\p}_{\!\xi^\Lambda} \right)
\right] g,
\label{MoyalT}
\ee
where $\eps$ is the refinement parameter to be related to $\mb$ in the next subsection.
It is easy to check that with respect to the Moyal product the relation \eqref{twitorus} gets deformed to
\be
\label{qtwitorus}
\cX_\gamma \star \cX_{\gamma'} =
(-y)^{\langle\gamma,\gamma'\rangle} \cX_{\gamma+\gamma'},
\qquad
y={\rm e}^{2\pi\I\epsilon}.
\ee
The function $\CU_\gamma$ in \eqref{qKStr}, which generates the quantum KS transformation,
is defined in terms the compact quantum dilogarithm $E_y(x)$, described in Appendix \ref{sec_qdilog}, as
\be
\CU_\gamma=\prod_{n\in\IZ} E_y\left(y^n
\cX_{\gamma}\right)^{\Omega_{n}(\gamma)},
\label{hviaU}
\ee
where $\Omega_{n}(\gamma)$ are the Laurent coefficients of the refined BPS indices
(to be distinguished from the multiplicities $\Omega_{[j]}(\gamma)$ defined in \eqref{defOmj})
\be
\Omega(\gamma,y) = \sum_{n\in\IZ} \Omega_{n}(\gamma) y^n.
\label{Omref}
\ee
Note that the product in \eqref{hviaU} is finite, since $\Omega_{n}(\gamma)$ vanishes for $|n|$ large enough.
Substituting~\eqref{hviaU} into \eqref{qKStr} and evaluating the star product explicitly, one finds
\begin{align}
\cX'_{\gamma'} ={}& \cX_{\gamma'}
\prod_{n\in\IZ}\left[ \frac{E_y\big(y^{n+\langle\gamma,\gamma'\rangle}  \cX_{\gamma}\big)}
{E_y\big(y^{n-\langle\gamma,\gamma'\rangle}  \cX_{\gamma}\big)}\right]^{\Omega_{n}(\gamma)}\nonumber
\\
={}& \cX_{\gamma'} \prod_{n\in\IZ}\prod_{k=0}^{|\langle\gamma,\gamma'\rangle|-1}
\big(1-y^{n+2k-|\langle\gamma,\gamma'\rangle|+1}\cX_{\gamma}
\big)^{\sign\langle\gamma,\gamma'\rangle \Omega_{n}(\gamma)} ,\label{transXg}
\end{align}
where in the second step we used the property \eqref{identEmany} of the quantum dilogarithm.
In the unrefined limit $y\to 1$, this transformation reduces to the classical symplectomorphism \eqref{KStr}, as it should.

The analysis of \cite{Alexandrov:2019rth} suggests that a similar non-commutative deformation
is induced by the refinement in full string theory as well. But in this case we also
need to understand how to extend $\hat\scU_\gamma$ to act on the full twistor space, including the coordinate $\talp$.
In other words, we need to extend the construction to the case of twisted
quantum tori.
To this end, we observe that the contact one-form \eqref{cXalp} on $\widetilde\CT_{\IC}$ arises by projectivizing
the symplectic form $\omega$ on a $\IC^\star$ bundle over $\widetilde\CT_{\IC}$ (which coincides
at least locally with the Swann bundle \cite{MR1096180} of the QK space $\CM$),
with Darboux coordinates $\big(\eta^I,\mu_I\big)$, $I=\flat,0,\dots,b_2$, such that \cite{Alexandrov:2010qdt}
\be
\omega=\sum_I \rd \eta^I \wedge \rd \mu_I  .
\label{symplform}
\ee
Indeed, under the identification
\be
\xi^{\Lambda}= \frac{\eta^\Lambda}{\eta^\flat} ,
\qquad
\txi_\Lambda=\mu_\Lambda ,
\qquad
\alpha= \mu_\flat ,
\ee
we have
\be
\omega = \rd\eta^\flat \wedge \big(\rd\mu_\flat+\xi^\Lambda\rd\txi_\Lambda\big)
+ \eta^\flat \rd \xi^\Lambda \wedge\rd \txi_\Lambda
= \rd\eta^\flat \wedge \CX + \eta^\flat \rd \CX .
\ee
Thus, we can define a star product on functions of $\xi^\Lambda$, $\txi_\Lambda$, $\alpha$
\big(or equivalently functions of $\xi^\Lambda$,~$\mu_I$\big) by viewing them as functions of $\big(\eta^I,\mu_I\big)$ which are
invariant under the $\IC^\times$ action $\eta^I\to \lambda \eta^I$,
and using the Moyal product on \smash{$\IC^{2b_2+4}$}, with a deformation parameter that we denote by \smash{$\eps^\flat$},
\be
f \star g = f \exp\left[ \frac{\I \eps^\flat}{2\pi} \sum_{I} \left(
\overleftarrow{\p}_{\!\eta^I}\overrightarrow{\p}_{\!\mu_I}
- \overleftarrow{\p}_{\!\mu_I}\overrightarrow{\p}_{\!\eta^I} \right)
\right] g.
\label{MoyalwT}
\ee
In general, the resulting Moyal product is not invariant under the $\IC^\times$ action $\eta^I\to \lambda \eta^I$,
unless this action also affects the deformation parameter via $\epsilon^\flat\to \lambda \epsilon^\flat$.
In other words, the result is not only a function of $\xi^\Lambda\equiv \eta^\Lambda/\eta^\flat$ and $\mu_I$ but also
depends on $\eps\equiv \eps^\flat/\eta^\flat$. Thus, the Moyal product \eqref{MoyalwT} defines
a non-commutative deformation of the product of functions
$f\big(\xi^\Lambda,\txi_\Lambda, \talp, \epsilon\big)$ on \smash{$\widetilde{T}_{\IC}\times \IC_\epsilon$},
which by construction preserves associativity.

It is easy to see that the corresponding Moyal bracket
\be
\{f,g\}_\star=\frac{\pi}{\I \eps}\(f \star g-g\star f\)
\label{Mbracket}
\ee
in the unrefined limit $\eps\to 0$ reproduces the contact bracket $\{ \cdot , \cdot \}_{0,0}$
introduced in \cite{Alexandrov:2008gh}.\footnote{In general, the contact bracket
is defined on sections of $\cO(2m)$ and $\cO(2n)$ bundles by
$
\smash{\{f, g\}_{m,n}=
\p_{\xi^\Lambda} f \p_{\txi_\Lambda} g }+
\(m -\xi^\Lambda \p_{\xi^\Lambda}\)f  \p_\alpha g
 - \p_{\xi^\Lambda} g \p_{\txi_\Lambda} f
-\(n -\xi^\Lambda \p_{\xi^\Lambda} \)g  \p_\alpha f$.
Arbitrary values of $m$ and $n$ can easily be incorporated into the above construction since
a section of $\cO(2m)$ bundle is described by a homogeneous function of degree~$m$ on \smash{$\widetilde{T}_{\IC}\times \IC_\epsilon$},
i.e., it is sufficient to postulate $f=\big(\eta^\flat\big)^m f\big(\eta^\Lambda/\eta^\flat,\mu_\Lambda,\mu_\flat,\eps^\flat/\eta^\flat\big)$
and evaluate the same star product~\eqref{MoyalwT}. However, in this work for our purposes it is sufficient to restrict to $m=0$.
}
Since the latter essentially encodes the contact structure, e.g., it generates classical contact transformations
via exponentiation $\exp\{ h, \cdot \}_{1,0}$ \cite{Alexandrov:2014mfa},
the star product \eqref{MoyalwT} can be thought as providing a definition of the refined contact structure.

\subsection{Wave functions and non-commutative wall-crossing}

Equipped with the star product, we can now construct a refined analogue of the dual partition function \eqref{symseries}.
To this end, we simply replace the usual product by the non-commutative one,\footnote{In the refined case,
we define \smash{$H^{\text{(ref)}}_k\big(\xi,\txi,\alpha\big)$} to be a function of $\alpha$ rather than $\talp$
because it is $\alpha$ that coincides with one of the Darboux coordinates for the symplectic form \eqref{symplform}
used to define the star product,
but one can always use \eqref{defalpha} to translate between the two variables.}
\be
H^{\text{(ref)}}_k\big(\xi,\txi,\alpha\big)=\sum_{\ell^\Lambda\in \frac{\IZ}{|k|\IZ} }\sum_{n^\Lambda\in\IZ+\ell^\Lambda/k}
{\rm e}^{2\pi\I k(\alpha+n^\Lambda \txi_\Lambda)}
\star \CH^{\text{(ref)}}_{k,\ell^\Lambda}\big(\xi^\Lambda-n^\Lambda\big).
\label{defZstar}
\ee
In fact, the star product can be evaluated explicitly using the property
\be
\label{keyprop}
{\rm e}^{2\pi\I( k\alpha+p^\Lambda \txi_\Lambda)}
\star f\big(\xi^\Lambda,\txi_\Lambda,\alpha, \eps\big)
= {\rm e}^{2\pi\I( k\alpha+p^\Lambda \txi_\Lambda)}
f\left(\frac{ \xi^\Lambda+\eps p^\Lambda}{1+\eps k} ,\txi_\Lambda,\alpha, \frac{\eps}{1+\eps k} \right).
\ee
In particular, it allows to see that the invariance under the Heisenberg group still holds.
Indeed, the star product changes the argument of \smash{$\CH^{\text{(ref)}}_{k,\ell^\Lambda}$}
to
\begin{gather*}
\frac{\xi^\Lambda+k\eps n^\Lambda}{1+k\eps}-n^\Lambda=\frac{\xi^\Lambda-n^\Lambda}{1+k\eps}
\end{gather*}
and thus the refined wave function is still a function of the difference \smash{$\xi^\Lambda-n^\Lambda$},
which ensures the invariance.

Our goal now is to determine the transformation property of the refined wave-function~\smash{$\CH^{\!\text{(ref)}}_{k,\ell}\!(\xi)$} under the quantum KS transformations \smash{$\hat\scU_\gamma$}
lifted to the twisted torus by means of~\eqref{MoyalwT}.
Denoting by \smash{$\hat\scV_\gamma$} the corresponding lift, we arrive at the following condition
\begin{gather}
\sum_{\ell^\Lambda,n^\Lambda}
{\rm e}^{2\pi\I k(\alpha+n^\Lambda \txi_\Lambda)}
\star \hat\scV_\gamma \big[\CH^{\text{(ref)}}_{k,\ell^\Lambda}(\xi-n)\big]\nonumber\\
\qquad
=\CU_\gamma \star\sum_{\ell^\Lambda,n^\Lambda}
{\rm e}^{2\pi\I k(\alpha+n^\Lambda \txi_\Lambda)}
\star \CH^{\text{(ref)}}_{k,\ell^\Lambda}(\xi-n)\star \CU_\gamma^{-1}.
\label{cond-trans}
\end{gather}
Restricting to the electric case $\gamma=(0,q_\Lambda)$, such that $\CU_\gamma$
becomes a function of $\xi^\Lambda$ and $\eps$ only and therefore commutes with the wave function,
and using \eqref{keyprop} to evaluate the star product, we obtain that the right-hand side of \eqref{cond-trans} is given by
\be
\sum_{\ell^\Lambda,n^\Lambda} {\rm e}^{2\pi\I k(\alpha+n^\Lambda \txi_\Lambda)}\star
\frac{\CU_\gamma\big(\frac{\xi-2k\eps n}{1-2k\eps};\frac{\eps}{1-2k\eps}\big)}{\CU_\gamma(\xi;\eps)}
\CH^{\text{(ref)}}_{k,\ell^\Lambda}(\xi-n).
\ee
Furthermore, using that
\be
{\rm e}^{-2\pi\I q_\Lambda\xi^\Lambda}= {\rm e}^{-2\pi\I q_\Lambda(\xi^\Lambda-n^\Lambda+\ell^\Lambda/k)},
\qquad
{\rm e}^{-2\pi\I q_\Lambda \frac{\xi^\Lambda-2k\eps n^\Lambda}{1-2k\eps}}
=
{\rm e}^{-2\pi\I q_\Lambda\( \frac{\xi^\Lambda-n^\Lambda}{1-2k\eps}+\frac{\ell^\Lambda}{k}\)},
\ee
it is easy to see that
the factor generated by the transformation depends on $\xi^\Lambda$ and $n^\Lambda$ only through their difference.
Thus, the condition \eqref{cond-trans} requires that the wave function should transform as
\be
\hat\scV_\gamma\big[\CH^{\text{(ref)}}_{k,\ell^\Lambda}(\xi)\big]
=\Upsilon_{k,\ell^\Lambda}(\xi)
 \CH^{\text{(ref)}}_{k,\ell^\Lambda}(\xi) ,
\qquad
\Upsilon_{k,\ell^\Lambda}(\xi)=\frac{\CU_\gamma\big(\frac{\xi}{1-2k\eps}+\frac{\ell}{k} ; \frac{\eps}{1-2k\eps}\big)}
{\CU_\gamma(\xi+\ell/k;\eps)} .
\label{trPsi-ref}
\ee

It turns out that the function $\Upsilon_{k,\ell^\Lambda}$ can be expressed through the Faddeev quantum dilogarithm
or its appropriate generalization, which are all described in Appendix \ref{sec_qdilog}.
For simplicity, let us first consider the unit multiplicity case, $\Omega_n(\gamma)=\delta_{n}$,
such that
\be
\CU_\gamma(\xi;\eps)=E_y\big({\rm e}^{-2\pi\I q_\Lambda\xi^\Lambda}\big)
\ee
with $y={\rm e}^{2\pi\I \epsilon}$.
We will distinguish between cases of positive and negative five-brane charge $k$
because they lead to different relations between the refinement parameters $\eps$ and $\mb$.
For $k>0$, we identify
\be
\eps=\frac{1}{2k}\big(1-\mb^{-2}\big),
\label{eps-b-pos}
\ee
such that $1-2k\eps =\mb^{- 2}$ and
\be
y = {\rm e}^{\pi \I/k} {\rm e}^{-\pi\I/( \mb^2k)}:=\ty_{\mb,k},
\qquad
{\rm e}^{ \frac{2\pi\I \eps}{1-2k\eps}}={\rm e}^{-\pi \I/k} {\rm e}^{\pi\I \mb^2/k}:=y_{\mb,k}.
\ee
which generalize the variables \eqref{defyb} to generic $k$ and satisfy $\ty_{\mb,k}=\by_{1/\bar\mb,k}$.
Thus, the function~$\Upsilon_{k,\ell^\Lambda}$ in \eqref{trPsi-ref} takes the form
\be
\Upsilon_{k,\ell^\Lambda}(\xi)=\frac{E_{y_{\mb,k}}\big({\rm e}^{-2\pi\I (\mb^2 q_\Lambda \xi^\Lambda +\ell/k)}\big)}
{E_{\ty_{\mb,k}}\big({\rm e}^{-2\pi\I (q_\Lambda \xi^\Lambda+\ell/k) }\big)}
 ,
\label{UspS}
\ee
where we denoted $\ell=q_\Lambda\ell^\Lambda$. Setting also $\snl=\sign(\ell)$ and using \eqref{identEmany},
the numerator can be rewritten as
\begin{gather}
E_{y_{\mb,k}}\big(y_{\mb,k}^{2\ell}  {\rm e}^{-2\pi\I \mb^2 (q_\Lambda \xi^\Lambda +\ell/k)}\big)\nonumber\\
\qquad= E_{y_{\mb,k}}\big({\rm e}^{-2\pi\I \mb^2 (q_\Lambda \xi^\Lambda +\ell/k)}\big) \prod_{j=0}^{|\ell|-1}
\big(1-y_{\mb,k}^{2\ell-\snl(2j+1)}  {\rm e}^{-2\pi\I \mb^2 (q_\Lambda\xi^\Lambda +\ell/k)}\big)^{\snl}.
\end{gather}
We can then use the property \eqref{Ekinverse} to get
\begin{align}
\Upsilon_{k,\ell^\Lambda}(\xi)&=
\prod_{j=0}^{k-1}\frac{E_{{\rm e}^{\pi\I \mb^2}}\bigl( -{\rm e}^{-2\pi\I \mb \Xi^+_{k,j}} \bigr)}
{E_{{\rm e}^{-\pi\I/ \mb^2}}\bigl(-{\rm e}^{-2\pi\I \mb^{-1} \Xi^+_{k,j}}\bigr)}
\prod_{j=0}^{|\ell|-1}\big(1- {\rm e}^{-2\pi\I( \mb^2 q_\Lambda\xi^\Lambda+\frac{\ell}{k}
+\frac{\snl}{k}(\mb^2-1)(j+\frac12))}\big)^{\snl}
\nonumber\\
&=
\prod_{j=0}^{k-1}\Phi_\mb^{-1}(-\I \Xi^+_{k,j})
\prod_{j=0}^{|\ell|-1}\big(1- {\rm e}^{-2\pi\I( \mb^2 q_\Lambda\xi^\Lambda+\frac{\ell}{k}
+\frac{\snl}{k}(\mb^2-1)(j+\frac12 ))}\big)^{\snl} ,
\label{Ups-choice1}
\end{align}
where
\be
\Xi^+_{k,j}=\mb q_\Lambda \xi^\Lambda+\frac{\mb\ell}{k}
-\frac{1}{k}\big(\mb-\mb^{-1}\big)\left(j-\frac{k-1}{2}\right).
\ee
For $\ell>0$ (hence $s=1$), \eqref{Ups-choice1} is recognized as
the function ${\bf A}_\hbar^{(k)}(t,\ell)$ defined in \cite[equation~(4.1)]{Alexandrov:2015xir}
evaluated at $\hbar=\mb^2$ and $t=-\I \mb^2 q_\Lambda \xi^\Lambda$.

For $k<0$, we replace the identification \eqref{eps-b-pos} by
\be
\eps=\frac{1}{2k}\big(1-\mb^2\big),
\label{eps-b-neg}
\ee
such that $1-2k\eps =\mb^2$ and
\be
y=y_{\mb,-k},
\qquad
{\rm e}^{\frac{2\pi\I \eps}{1-2k\eps}}=\ty_{\mb,-k}.
\ee
Repeating the same steps as above, one arrives at the following
representation for the function~$\Upsilon_{k,\ell^\Lambda}$ \eqref{trPsi-ref}:
\begin{gather}
\Upsilon_{k,\ell^\Lambda}(\xi)
=
\prod_{j=0}^{|k|-1}\Phi_\mb(-\I \Xi^-_{k,j})
\prod_{j=0}^{|\ell|-1}\big(1- {\rm e}^{-2\pi\I( \mb^{-2} q_\Lambda\xi^\Lambda+\frac{\ell}{k}
+\frac{\snl}{k}(\mb^{-2}-1)(j+\frac12 ))}\big)^{\snl} ,
\label{Ups-choice2}
\end{gather}
where now
\be
\Xi^-_{k,j}=\mb^{-1} q_\Lambda \xi^\Lambda+\frac{\ell}{\mb k}
-\frac{1}{|k|}\big(\mb-\mb^{-1}\big)\left(j-\frac{|k|-1}{2}\right).
\ee

For BPS rays carrying general refined BPS indices $\Omega_n(\gamma)$, the wall-crossing transformation
can be expressed through a generalization \smash{$\Phi^{[j]}_{\mb,k}(z)$} of the Faddeev quantum dilogarithm defined in \eqref{defPhibjk},
which also has a representation as a product of the usual quantum dilogarithms with shifted arguments.
Again, proceeding as above, it is straightforward to show that
\begin{align}
\Upsilon_{k,\ell^\Lambda}(\xi)={}&\prod_j\left[
\prod_{i=0}^{|k|-1}\(\Phi^{[j]}_{\mb,k}(-\I \Xi^{\snk}_{k,i})\)^{\snk}\right.\nonumber\\
&\left.\times
\prod_{m=-j}^j
\prod_{i=0}^{|\ell|-1}\big(1- {\rm e}^{-2\pi\I( \mb^{2\snk} q_\Lambda\xi^\Lambda+\frac{\ell}{k}
+\frac{\snl}{k}(\mb^{2\snk}-1)(i-m+\frac12 ))}\big)^{-\snl}
\right]^{-\Omega_{[j]}} ,
\label{Ups-choice1j}
\end{align}
where $\snk=\sign(k)$.
For $k=1$, in which case $\ell=0$, this formula reduces to
\begin{gather}
\Upsilon_{1}(\xi) =
\prod_j\big[ \Phi_{\mb}^{[j]}\bigl(-\I \mb q_\Lambda\xi^\Lambda\bigr)\big]^{-\Omega_{[j]}}.
\end{gather}
This result is to be compared with \eqref{mfrakmultiLR} where $\CA=2\pi q_a t^a+4\pi^2 \I q_0$.
It is easy to see that the factor appearing in \eqref{mfrakmultiLR} coincides with $\Upsilon_{1}(\xi)$ provided
one relates the variables as in \eqref{ident-arg} with $\xi^\Lambda$ replaced by $\mb \xi^\Lambda$, i.e.,
\be
t^a=2\pi\I \frac{\xi^a}{\xi^0} ,
\qquad
g_s=\frac{2\pi}{\mb\xi^0} .
\label{ident-arg-b}
\ee
Note that the last relation is equivalent to
\be
\xi^0=\frac{2\pi}{\eps_1} ,
\label{ident-xi0eps}
\ee
where $\eps_1$ is one the deformation parameters of the $\Omega$-background, see \eqref{ep-beta}.
The match of the Stokes factors suggests that the wave function \smash{$\CH^{\text{(ref)}}_{1}$} is
equal to the refined topological string up to a $\mb$-dependent constant,
\be
\CH^{\text{(ref)}}_{1}(\xi)\sim \CZ(\bft;g_s,\mb),
\label{H-Z1}
\ee
with the parameters identified as in \eqref{ident-arg-b}. This generalizes a similar relation in the unrefined case \cite{APP}.

Note that if we had chosen the opposite ordering in the definition of the refined dual partition function \eqref{defZstar},
the results for the cases of positive and negative $k$ would effectively be swapped.
Indeed, exchanging the ordering of factors in \eqref{keyprop}
leads to the flip of signs in front of $k$ and $p^\Lambda$ in the arguments of
the function $f$. As a result, denoting by tilde the quantities corresponding to the opposite ordering, one has
\be
\tUps_{k,\ell^\Lambda}(\xi)=\Upsilon_{-k,-\ell^\Lambda}^{-1}(\xi).
\ee
In particular, using \eqref{Ups-choice1j}, one finds that
\be
\tUps_{1}(\xi) =
\prod_j\big[ \Phi_{\mb}^{[j]}\bigl(-\I \mb^{-1} q_\Lambda\xi^\Lambda\bigr)\big]^{-\Omega_{[j]}},
\ee
which in turn implies the identification
\be
\tilde\CH^{\text{(ref)}}_{1}(\xi)\sim \CZ\left(2\pi\I \frac{\xi^a}{\xi^0} ; \frac{2\pi\mb}{\xi^0} ,\mb\right).
\label{H-Z2}
\ee
Note that in this case the relation \eqref{ident-xi0eps} is replaced by
\be
\xi^0=-\frac{2\pi}{\eps_2} .
\label{ident-xi0eps2}
\ee

We observe that the effect of the refinement is to introduce a factor of $\mb$ in the relation between $\xi^0$
and the topological string coupling $g_s$. It spoils the symmetry $\mb\leftrightarrow \mb^{-1}$,
unless one simultaneously changes the ordering in the definition of the refined dual partition function.
Of course, this factor could be absorbed in the definition of the Darboux coordinates $\xi^\Lambda$,
but the price to pay is a modification of the quasi-periodicity conditions \eqref{HeisenZ} and \eqref{thetaline},
and it would reappear anyway in the quantization condition of charges.

\appendix

\section{Quantum dilogarithms}
\label{sec_qdilog}

In this Appendix, we introduce several versions of the quantum dilogarithm function which play a role in this work.

The standard (sometimes called compact)
quantum dilogarithm $E_y(x)$ is defined for ${x,y\in\IC}$, $|y|<1$
as\footnote{There are different
conventions in the literature, e.g., \cite{Barbieri:2019yya,Chuang:2022uey}
define the quantum dilogarithm as $\mathbb{E}_q(x)=(x,q)_{\infty}$,
which is related to our definition by $E_y(x)=[\mathbb{E}_{y^2}(xy)]^{-1}$.}
\be
\label{defEy}
E_y(x) := \exp\left[ \sum_{k=1}^{\infty} \frac{(x y)^k}{k\big(1-y^{2k}\big)} \right] = \prod_{n=0}^{\infty}
\big(1-x y^{2n+1}\big)^{-1} = \big(xy;y^2\big)_{\infty}^{-1},
\ee
where
$(a;q)_n:=\prod_{k=0}^{n-1} \big(1-a q^k\big)$ is the $q$-Pochhammer symbol.
The quantum dilogarithm satisfies the following properties:
\begin{gather}
E_y\big(xy^2 \big)=(1-x y)E_y(x),
\label{identE} \\
E_{y^{1/k}}(x)=\prod_{j=0}^{k-1} E_y\( y^{\frac{2j+1}{k}-1}x\),
\qquad k\in \IN.
\label{Ekinverse}
\end{gather}
The first property has the obvious generalization
\be
E_y\big(y^{2\ell} x\big) =E_y(x) \prod_{j=0}^{\ell-1} \big(1-x y^{2j+1}\big),
\qquad
\ell\in \IN.
\label{identEmany}
\ee

A different version of the quantum dilogarithm (sometimes called non-compact)
was introduced by Faddeev \cite{Faddeev}, through the contour integral
\be
\Phi_\mb(z) := \exp\left( \int_{\IR+\I \epsilon} \frac{{\rm e}^{-2\I z v}}{4\sinh(v \mb) \sinh(v/\mb)} \frac{\rd v}{v} \right)
\label{defPhib}
\ee
over the real line, circumventing the pole at $v=0$ by deviating into the upper half plane. This
integral converges for $\Re \mb\neq 0$, $|\Im z|<|\Im c_{\mb}|$ with $c_{\mb}:=\frac{\I}{2}\big(\mb+\mb^{-1}\big)$.
It possesses several beautiful properties listed, for example, in \cite[Section~A.2]{Alexandrov:2015xir}.
Here we mention the quasi-periodicity
\be
\Phi_\mb\big(x-\I \mb^{\pm 1}/2\big)=\big(1+{\rm e}^{2\pi \mb^{\pm 1} x}\big)\Phi_\mb\big(x+\I \mb^{\pm 1}/2\big),
\label{shiftPhi}
\ee
the classical limit $\mb\to 0$ \cite[Section~13.4]{EllegaardAndersen:2011vps}
\be
\Phi_\mb(x)= \exp\left(\frac{\Li_2\bigl(-{\rm e}^{2\pi \mb x}\bigr)}{2\pi\I \mb^2}\right)
\( 1+ \cO\big(\mb^2\big) \) ,
\label{classPhi}
\ee
the special value at $\mb=1$
\be
\Phi_1(x)=\exp\[\frac{\I}{2\pi}\(\Li_2\big({\rm e}^{2\pi x}\big)+2\pi x\log\big(1-{\rm e}^{2\pi x}\big)\)\],
\label{onePhi}
\ee
and the special value at $\mb=\sqrt{2}$
\be
\label{bsq2}
\Phi_{\sqrt{2}} \left( {x \over {\sqrt{2}}} \right)=
\exp\[\frac{\I}{4\pi}\(\Li_2\bigl(-{\rm e}^{2\pi x}\bigr)+2\pi x\log\big(1+{\rm e}^{2\pi x}\big) +2 \pi \tan^{-1} \big(\re^{\pi x}\big)\)\],
\ee
which follows from the more general results obtained in \cite{garkas}.

Evaluating the integral in \eqref{defPhib} by residues, one finds that
\be
\log\Phi_\mb(z) = \sum_{\ell=1}^{\infty} \frac{(-1)^\ell}{2\I \ell}
\left( \frac{{\rm e}^{2\pi \ell z \mb}}{\sin\big(\ell\pi \mb^2\big)} + \frac{{\rm e}^{2\pi \ell z/\mb}}{\sin\big(\ell\pi/\mb^2\big)} \right),
\label{Phibsum}
\ee
which allows to establish the following relation between the two versions of quantum dilogarithm~\cite[Section~A]{Faddeev:2000if}, valid whenever $\Im \big(b^2\big)>0$,
\be
\label{PhiE}
\Phi_\mb(z) = \frac{E_{{\rm e}^{-\I\pi/\mb^2}}\bigl( - {\rm e}^{2\pi z/\mb} \bigr)}
{E_{{\rm e}^{\I\pi \mb^2}}\bigl( - {\rm e}^{2\pi z \mb} \bigr)}
=\frac{E_{\ty_\mb}\big( {\rm e}^{2\pi z/\mb} \big)}
{E_{y_\mb}\big( {\rm e}^{2\pi z \mb} \big)}
 ,
\ee
where in the second representation we used the variables defined in \eqref{defyb}.

Let us now introduce a generalization of the compact quantum dilogarithm,
which appeared in \cite[equation~(2.12)]{Dimofte:2009bv} and depends on an additional label $j\in \IZ/2$,
\be
\label{defEyj}
E_{y}^{[j]}(x) = \exp\left[ \sum_{k=1}^{\infty} \frac{(x y)^k \chi_j\big(y^{k}\big)}{k\big(1-y^{2k}\big)} \right]
=\prod_{m=-j}^j \big(xy^{2m+1};y^2\big)_{\infty}^{-1},
\ee
where $\chi_j(y)$ is the character \eqref{defcharacter} of the ${\rm SU}(2)$ representation of spin $j$.
It is easy to see that%
\be
\label{Eyjm}
E_{y}^{[j]}(x) = \prod_{m=-j}^j E_y\big(xy^{2m}\big),
\ee
where the product runs over half-integers $m$ such that $m-j$ is integer.
Following \eqref{PhiE}, we then define the corresponding generalization of the Faddeev quantum dilogarithm
\be
\Phi_{\mb}^{[j]}(z) = \frac{E_{\ty_\mb}^{[j]}\big( {\rm e}^{2\pi z/\mb} \big)}
{E_{y_\mb}^{[j]}\big( {\rm e}^{2\pi z \mb} \big)} .
\label{defPhibj}
\ee
From this definition and \eqref{defEyj}, it follows that this new function can be written in one of the following forms
\begin{align}
\Phi_{\mb}^{[j]} (z) ={}& \exp\[ \sum_{\ell=1}^{\infty} \frac{(-1)^\ell}{2\I \ell}
\left( \frac{\chi_j\bigl(-{\rm e}^{\pi\I \ell\mb^2}\bigr) {\rm e}^{2\pi \ell z \mb}}{\sin\big(\ell\pi \mb^2\big)}
+ \frac{\chi_j\bigl(-{\rm e}^{-\pi\I \ell/\mb^2}\bigr) {\rm e}^{2\pi \ell z/\mb}}{\sin\big(\ell\pi/\mb^2\big)} \right)\]\nonumber
\\
={}&
\exp\left( \int_{\IR+\I \epsilon} \frac{\chi_j\big({\rm e}^{ (\mb-\mb^{-1}) v}\big)
{\rm e}^{-2\I z v}}{4\sinh(v \mb) \sinh(v/\mb)} \frac{\rd v}{v} \right)\nonumber
\\
={}&  \prod_{m=-j}^j \Phi_\mb\big(z+\ri m\big(\mb-\mb^{-1}\big)\big)\label{reprPhibj}
\end{align}
in accordance with \eqref{Eyjm}.

Finally, we incorporate one additional integer parameter $k$ corresponding to the five-brane charge.
To this end, we define
\be
\label{Eyjmk}
E^{[j]}_{y,k}(x) = \prod_{m=-j}^j E_y\big(xy^{2m/k}\big)
\ee
and
\be
\Phi^{[j]}_{\mb,k}(z) := \frac{E^{[j]}_{\ty_\mb,k}\big({\rm e}^{2\pi z/\mb} \big)}
{E^{[j]}_{y_\mb,k}\big({\rm e}^{2\pi z \mb} \big)}
= \prod_{m=-j}^j \Phi_\mb\left(z+\frac{\ri m}{k} \big(\mb-\mb^{-1}\big)\right) .
\label{defPhibjk}
\ee

\subsection*{Acknowledgements}

The authors would like to thank Murad Alim, Tom Bridgeland, Alba Grassi,
Jie Gu and Joerg Teschner for valuable discussions. We would also like to thank the anonymous referees for their
useful and detailed comments, which helped in improving the presentation of our results.
The research of MM is supported in part by the ERC-SyG project
``Recursive and Exact New Quantum Theory" (ReNewQuantum), which
received funding from the European Research Council (ERC) under the European
Union's Horizon 2020 research and innovation program, grant agreement No. 810573.
The research of BP is supported by Agence Nationale de la Recherche under contract number ANR-21-CE31-0021.
SA and BP would like to thank the Isaac Newton Institute for Mathematical Sciences in Cambridge
(supported by EPSRC grant EP/R014604/1), for hospitality
during the programme {\it Black holes: bridges between number theory and holographic quantum information}
where work on this project was undertaken.

\pdfbookmark[1]{References}{ref}
\LastPageEnding


\begin{thebibliography}{99}
\footnotesize\itemsep=0pt

\bibitem{Aganagic:2006wq}
Aganagic M., Bouchard V., Klemm A., Topological strings and (almost) modular
 forms, \href{https://doi.org/10.1007/s00220-007-0383-3}{\textit{Comm. Math. Phys.}} \textbf{277} (2008), 771--819,
 \href{https://arxiv.org/abs/hep-th/0607100}{arXiv:hep-th/0607100}.

\bibitem{adkmv}
Aganagic M., Dijkgraaf R., Klemm A., Mari\~no M., Vafa C., Topological strings
 and integrable hierarchies, \href{https://doi.org/10.1007/s00220-005-1448-9}{\textit{Comm. Math. Phys.}} \textbf{261} (2006),
 451--516, \href{https://arxiv.org/abs/hep-th/0312085}{arXiv:hep-th/0312085}.

\bibitem{akmv}
Aganagic M., Klemm A., Mari\~no M., Vafa C., The topological vertex,
 \href{https://doi.org/10.1007/s00220-004-1162-z}{\textit{Comm. Math. Phys.}} \textbf{254} (2005), 425--478,
 \href{https://arxiv.org/abs/hep-th/0305132}{arXiv:hep-th/0305132}.

\bibitem{Alexandrov:2009zh}
Alexandrov S., D-instantons and twistors: some exact results,
 \href{https://doi.org/10.1088/1751-8113/42/33/335402}{\textit{J.~Phys.~A}} \textbf{42} (2009), 335402, 26~pages, \href{https://arxiv.org/abs/0902.2761}{arXiv:0902.2761}.

\bibitem{Alexandrov:2014rca}
Alexandrov S., Banerjee S., Dualities and fivebrane instantons, \href{https://doi.org/10.1007/JHEP11(2014)040}{\textit{J.~High
 Energy Phys.}} \textbf{2014} (2014), no.~11, 040, 39~pages,
 \href{https://arxiv.org/abs/1405.0291}{arXiv:1405.0291}.

\bibitem{Alexandrov:2014mfa}
Alexandrov S., Banerjee S., Fivebrane instantons in {C}alabi--{Y}au
 compactifications, \href{https://doi.org/10.1103/PhysRevD.90.041902}{\textit{Phys. Rev.~D}} \textbf{90} (2014), 041902, 5~pages,
 \href{https://arxiv.org/abs/1403.1265}{arXiv:1403.1265}.

\bibitem{Alexandrov:2023hiv}
Alexandrov S., Bendriss K., Hypermultiplet metric and {NS}5-instantons,
 \href{https://doi.org/10.1007/jhep01(2024)140}{\textit{J.~High Energy Phys.}} \textbf{2024} (2024), no.~1, 140, 45~pages,
 \href{https://arxiv.org/abs/2309.14440}{arXiv:2309.14440}.

\bibitem{Alexandrov:2019rth}
Alexandrov S., Manschot J., Pioline B., S-duality and refined {BPS} indices,
  \href{https://doi.org/10.1007/s00220-020-03854-6}{\textit{Comm. Math. Phys.}} \textbf{380} (2020), 755--810,
 \href{https://arxiv.org/abs/1910.03098}{arXiv:1910.03098}.

\bibitem{APP}
Alexandrov S., Persson D., Pioline B., Fivebrane instantons, topological wave
 functions and hypermultiplet moduli spaces, \href{https://doi.org/10.1007/JHEP03(2011)111}{\textit{J.~High Energy Phys.}}
 \textbf{2011} (2011), no.~3, 111, 73~pages, \href{https://arxiv.org/abs/1010.5792}{arXiv:1010.5792}.

\bibitem{Alexandrov:2010np}
Alexandrov S., Persson D., Pioline B., On the topology of the hypermultiplet
 moduli space in type {II}/{CY} string vacua, \href{https://doi.org/10.1103/PhysRevD.83.026001}{\textit{Phys. Rev.~D}}
 \textbf{83} (2011), 026001, 5~pages, \href{https://arxiv.org/abs/1009.3026}{arXiv:1009.3026}.

\bibitem{Alexandrov:2011ac}
Alexandrov S., Persson D., Pioline B., Wall-crossing, {R}ogers dilogarithm, and
 the {QK}/{HK} correspondence, \href{https://doi.org/10.1007/JHEP12(2011)027}{\textit{J.~High Energy Phys.}} \textbf{2011}
 (2011), no.~12, 027, 64~pages, \href{https://arxiv.org/abs/1110.0466}{arXiv:1110.0466}.

\bibitem{Alexandrov:2015xir}
Alexandrov S., Pioline B., Theta series, wall-crossing and quantum dilogarithm
 identities, \href{https://doi.org/10.1007/s11005-016-0857-3}{\textit{Lett. Math. Phys.}} \textbf{106} (2016), 1037--1066,
 \href{https://arxiv.org/abs/1511.02892}{arXiv:1511.02892}.

\bibitem{Alexandrov:2021wxu}
Alexandrov S., Pioline B., Heavenly metrics, {BPS} indices and twistors,
  \href{https://doi.org/10.1007/s11005-021-01455-5}{\textit{Lett. Math. Phys.}} \textbf{111} (2021), 116, 41~pages,
 \href{https://arxiv.org/abs/2104.10540}{arXiv:2104.10540}.

\bibitem{Alexandrov:2008gh}
Alexandrov S., Pioline B., Saueressig F., Vandoren S., D-instantons and
 twistors, \href{https://doi.org/10.1088/1126-6708/2009/03/044}{\textit{J.~High Energy Phys.}} \textbf{2009} (2009), no.~3, 044,
 40~pages, \href{https://arxiv.org/abs/0812.4219}{arXiv:0812.4219}.

\bibitem{Alexandrov:2010qdt}
Alexandrov S., Pioline B., Saueressig F., Vandoren S., Linear perturbations of
 quaternionic metrics, \href{https://doi.org/10.1007/s00220-010-1022-y}{\textit{Comm. Math. Phys.}} \textbf{296} (2010),
 353--403, \href{https://arxiv.org/abs/0810.1675}{arXiv:0810.1675}.

\bibitem{Alim:2022oll}
Alim M., Hollands L., Tulli I., Quantum curves, resurgence and exact {WKB},
  \href{https://doi.org/10.3842/SIGMA.2023.009}{\textit{SIGMA}} \textbf{19} (2023), 009, 82~pages, \href{https://arxiv.org/abs/2203.08249}{arXiv:2203.08249}.

\bibitem{al}
Alim M., L\"ange J.D., Polynomial structure of the (open) topological string
 partition function,  \href{https://doi.org/10.1088/1126-6708/2007/10/045}{\textit{J.~High Energy Phys.}} \textbf{2007} (2007),
 no.~10, 045, 13~pages, \href{https://arxiv.org/abs/0708.2886}{arXiv:0708.2886}.

\bibitem{astt}
Alim M., Saha A., Teschner J., Tulli I., Mathematical structures of
 non-perturbative topological string theory: from {GW} to {DT} invariants,
  \href{https://doi.org/10.1007/s00220-022-04571-y}{\textit{Comm. Math. Phys.}} \textbf{399} (2023), 1039--1101,
 \href{https://arxiv.org/abs/2109.06878}{arXiv:2109.06878}.

\bibitem{EllegaardAndersen:2011vps}
Andersen J.E., Kashaev R., A {TQFT} from quantum {T}eichm\"uller theory,
 \href{https://doi.org/10.1007/s00220-014-2073-2}{\textit{Comm. Math. Phys.}} \textbf{330} (2014), 887--934, \href{https://arxiv.org/abs/1109.6295}{arXiv:1109.6295}.

\bibitem{Antoniadis:2013bja}
Antoniadis I., Florakis I., Hohenegger S., Narain K.S., Zein~Assi A.,
 Worldsheet realization of the refined topological string, \href{https://doi.org/10.1016/j.nuclphysb.2013.07.004}{\textit{Nuclear
 Phys.~B}} \textbf{875} (2013), 101--133, \href{https://arxiv.org/abs/1302.6993}{arXiv:1302.6993}.

\bibitem{Barbieri:2019yya}
Barbieri A., Bridgeland T., Stoppa J., A quantized {R}iemann--{H}ilbert problem
 in {D}onaldson--{T}homas theory, \href{https://doi.org/10.1093/imrn/rnaa294}{\textit{Int. Math. Res. Not.}} \textbf{2022}
 (2022), 3417--3456, \href{https://arxiv.org/abs/1905.00748}{arXiv:1905.00748}.

\bibitem{bcov}
Bershadsky M., Cecotti S., Ooguri H., Vafa C., Kodaira--{S}pencer theory of
 gravity and exact results for quantum string amplitudes,  \href{https://doi.org/10.1007/BF02099774}{\textit{Comm. Math.
 Phys.}} \textbf{165} (1994), 311--427, \href{https://arxiv.org/abs/hep-th/9309140}{arXiv:hep-th/9309140}.

\bibitem{bkmp}
Bouchard V., Klemm A., Mari\~no M., Pasquetti S., Remodeling the {B}-model,
  \href{https://doi.org/10.1007/s00220-008-0620-4}{\textit{Comm. Math. Phys.}} \textbf{287} (2009), 117--178, \href{https://arxiv.org/abs/0709.1453}{arXiv:0709.1453}.

\bibitem{Bri-19}
Bridgeland T., Riemann--{H}ilbert problems from {D}onaldson--{T}homas theory,
  \href{https://doi.org/10.1007/s00222-018-0843-8}{\textit{Invent. Math.}} \textbf{216} (2019), 69--124, \href{https://arxiv.org/abs/1611.03697}{arXiv:1611.03697}.

\bibitem{Bri-23}
Bridgeland T., Tau functions from {J}oyce structures, \href{https://arxiv.org/abs/2303.07061}{arXiv:2303.07061}.

\bibitem{Bridgeland:2020zjh}
Bridgeland T., Strachan I.A.B., Complex hyperk\"ahler structures defined by
 {D}onaldson--{T}homas invariants, \href{https://doi.org/10.1007/s11005-021-01388-z}{\textit{Lett. Math. Phys.}} \textbf{111}
 (2021), 54, 24~pages, \href{https://arxiv.org/abs/2006.13059}{arXiv:2006.13059}.

\bibitem{Cecotti:2014wea}
Cecotti S., Neitzke A., Vafa C., Twistorial topological strings and a~{$tt^*$}
 geometry for~{$\mathcal{N}=2$} theories in~{$4d$}, \href{https://doi.org/10.4310/ATMP.2016.v20.n2.a1}{\textit{Adv. Theor. Math.
 Phys.}} \textbf{20} (2016), 193--312, \href{https://arxiv.org/abs/1412.4793}{arXiv:1412.4793}.

\bibitem{ckk}
Choi J., Katz S., Klemm A., The refined {BPS} index from stable pair
 invariants, \href{https://doi.org/10.1007/s00220-014-1978-0}{\textit{Comm. Math. Phys.}} \textbf{328} (2014), 903--954,
 \href{https://arxiv.org/abs/1210.4403}{arXiv:1210.4403}.

\bibitem{Chuang:2022uey}
Chuang W., Quantum {R}iemann--{H}ilbert problems for the resolved conifold,
  \href{https://doi.org/10.1016/j.geomphys.2023.104860}{\textit{J.~Geom. Phys.}} \textbf{190} (2023), 104860, 16~pages,
 \href{https://arxiv.org/abs/2203.00294}{arXiv:2203.00294}.

\bibitem{Codesido:2015dia}
Codesido S., Grassi A., Mari\~no M., Spectral theory and mirror curves of
 higher genus, \href{https://doi.org/10.1007/s00023-016-0525-2}{\textit{Ann. Henri Poincar\'e}} \textbf{18} (2017), 559--622,
 \href{https://arxiv.org/abs/1507.02096}{arXiv:1507.02096}.

\bibitem{CLT}
Coman I., Longhi P., Teschner J., From quantum curves to topological string
 partition functions~{II}, \href{https://arxiv.org/abs/2004.04585}{arXiv:2004.04585}.

\bibitem{cpt}
Coman I., Pomoni E., Teschner J., From quantum curves to topological string
 partition functions, \href{https://doi.org/10.1007/s00220-022-04579-4}{\textit{Comm. Math. Phys.}} \textbf{399} (2023),
 1501--1548, \href{https://arxiv.org/abs/1811.01978}{arXiv:1811.01978}.

\bibitem{cesv2}
Couso-Santamar\'{\i}a R., Edelstein J.D., Schiappa R., Vonk M., Resurgent
 transseries and the holomorphic anomaly: nonperturbative closed strings in
 local~{$\mathbb{CP}^2$}, \href{https://doi.org/10.1007/s00220-015-2358-0}{\textit{Comm. Math. Phys.}} \textbf{338} (2015),
 285--346, \href{https://arxiv.org/abs/1407.4821}{arXiv:1407.4821}.

\bibitem{cesv1}
Couso-Santamar\'{\i}a R., Edelstein J.D., Schiappa R., Vonk M., Resurgent
 transseries and the holomorphic anomaly, \href{https://doi.org/10.1007/s00023-015-0407-z}{\textit{Ann. Henri Poincar\'e}}
 \textbf{17} (2016), 331--399, \href{https://arxiv.org/abs/1308.1695}{arXiv:1308.1695}.

\bibitem{cms}
Couso-Santamar\'{\i}a R., Mari\~no M., Schiappa R., Resurgence matches
 quantization, \href{https://doi.org/10.1088/1751-8121/aa5e01}{\textit{J.~Phys.~A}} \textbf{50} (2017), 145402, 34~pages,
 \href{https://arxiv.org/abs/1610.06782}{arXiv:1610.06782}.

\bibitem{DDP93}
Delabaere E., Dillinger H., Pham F., R\'esurgence de {V}oros et p\'eriodes des
 courbes hyperelliptiques, \href{https://doi.org/10.5802/aif.1326}{\textit{Ann. Inst. Fourier (Grenoble)}} \textbf{43}
 (1993), 163--199.

\bibitem{dhsv}
Dijkgraaf R., Hollands L., Su{\l}kowski P., Vafa C., Supersymmetric gauge
 theories, intersecting branes and free fermions, \href{https://doi.org/10.1088/1126-6708/2008/02/106}{\textit{J.~High Energy
 Phys.}} \textbf{2008} (2008), no.~2, 106, 57~pages, \href{https://arxiv.org/abs/0709.4446}{arXiv:0709.4446}.

\bibitem{Dimofte:2009bv}
Dimofte T., Gukov S., Refined, motivic, and quantum, \href{https://doi.org/10.1007/s11005-009-0357-9}{\textit{Lett. Math. Phys.}}
 \textbf{91} (2010), 1--27, \href{https://arxiv.org/abs/0904.1420}{arXiv:0904.1420}.

\bibitem{dmp-np}
Drukker N., Mari\~no M., Putrov P., Nonperturbative aspects of {ABJM} theory,
 \href{https://doi.org/10.1007/JHEP11(2011)141}{\textit{J.~High Energy Phys.}} \textbf{2011} (2011), no.~11, 141, 29~pages,
 \href{https://arxiv.org/abs/1103.4844}{arXiv:1103.4844}.

\bibitem{Eynard:2021sxg}
Eynard B., Garcia-Failde E., Marchal O., Orantin N., Quantization of classical
 spectral curves via topological recursion, \href{https://doi.org/10.1007/s00220-024-04997-6}{\textit{Comm. Math. Phys.}}
 \textbf{405} (2024), 116, 118~pages, \href{https://arxiv.org/abs/2106.04339}{arXiv:2106.04339}.

\bibitem{em}
Eynard B., Mari\~no M., A holomorphic and background independent partition
 function for matrix models and topological strings, \href{https://doi.org/10.1016/j.geomphys.2010.11.012}{\textit{J.~Geom. Phys.}}
 \textbf{61} (2011), 1181--1202, \href{https://arxiv.org/abs/0810.4273}{arXiv:0810.4273}.

\bibitem{eo}
Eynard B., Orantin N., Invariants of algebraic curves and topological
 expansion, \href{https://doi.org/10.4310/CNTP.2007.v1.n2.a4}{\textit{Commun. Number Theory Phys.}} \textbf{1} (2007), 347--452,
 \href{https://arxiv.org/abs/math-ph/0702045}{arXiv:math-ph/0702045}.

\bibitem{Faddeev}
Faddeev L.D., Discrete {H}eisenberg--{W}eyl group and modular group,
 \href{https://doi.org/10.1007/BF01872779}{\textit{Lett. Math. Phys.}} \textbf{34} (1995), 249--254,
 \href{https://arxiv.org/abs/hep-th/9504111}{arXiv:hep-th/9504111}.

\bibitem{Faddeev:2000if}
Faddeev L.D., Kashaev R.M., Volkov A.Yu., Strongly coupled quantum discrete
 {L}iouville theory.~{I}. {A}lgebraic approach and duality, \href{https://doi.org/10.1007/s002200100412}{\textit{Comm.
 Math. Phys.}} \textbf{219} (2001), 199--219, \href{https://arxiv.org/abs/hep-th/0006156}{arXiv:hep-th/0006156}.

\bibitem{Ferrara:1989ik}
Ferrara S., Sabharwal S., Quaternionic manifolds for type~{${\rm II}$}
 superstring vacua of {C}alabi--{Y}au spaces, \href{https://doi.org/10.1016/0550-3213(90)90097-W}{\textit{Nuclear Phys.~B}}
 \textbf{332} (1990), 317--332.

\bibitem{Gaiotto:2008cd}
Gaiotto D., Moore G.W., Neitzke A., Four-dimensional wall-crossing via
 three-dimensional field theory, \href{https://doi.org/10.1007/s00220-010-1071-2}{\textit{Comm. Math. Phys.}} \textbf{299}
 (2010), 163--224, \href{https://arxiv.org/abs/0807.4723}{arXiv:0807.4723}.

\bibitem{Gaiotto:2010be}
Gaiotto D., Moore G.W., Neitzke A., Framed {BPS} states, \href{https://doi.org/10.4310/ATMP.2013.v17.n2.a1}{\textit{Adv. Theor.
 Math. Phys.}} \textbf{17} (2013), 241--397, \href{https://arxiv.org/abs/1006.0146}{arXiv:1006.0146}.

\bibitem{GIL-12}
Gamayun O., Iorgov N., Lisovyy O., Conformal field theory of {P}ainlev\'e~{VI},
 \href{https://doi.org/10.1007/JHEP10(2012)038}{\textit{J.~High Energy Phys.}} \textbf{2012} (2012), no.~10, 038, 24~pages,
 \href{https://arxiv.org/abs/1207.0787}{arXiv:1207.0787}.

\bibitem{GIL-13}
Gamayun O., Iorgov N., Lisovyy O., How instanton combinatorics solves
 {P}ainlev\'e~{VI}, {V} and~{III}s, \href{https://doi.org/10.1088/1751-8113/46/33/335203}{\textit{J.~Phys.~A}} \textbf{46} (2013),
 335203, 29~pages, \href{https://arxiv.org/abs/1302.1832}{arXiv:1302.1832}.

\bibitem{garkas}
Garoufalidis S., Kashaev R., Evaluation of state integrals at rational points,
 \href{https://doi.org/10.4310/CNTP.2015.v9.n3.a3}{\textit{Commun. Number Theory Phys.}} \textbf{9} (2015), 549--582,
 \href{https://arxiv.org/abs/1411.6062}{arXiv:1411.6062}.

\bibitem{gv-conifold}
Ghoshal D., Vafa C., {$c=1$} string as the topological theory of the conifold,
 \href{https://doi.org/10.1016/0550-3213(95)00408-K}{\textit{Nuclear Phys.~B}} \textbf{453} (1995), 121--128,
 \href{https://arxiv.org/abs/hep-th/9506122}{arXiv:hep-th/9506122}.

\bibitem{gv}
Gopakumar R., Vafa C., M-theory and topological strings.~{II},
 \href{https://arxiv.org/abs/hep-th/9812127}{arXiv:hep-th/9812127}.

\bibitem{Grassi:2022zuk}
Grassi A., Hao Q., Neitzke A., Exponential networks, {WKB} and topological
 string, \href{https://doi.org/10.3842/SIGMA.2023.064}{\textit{SIGMA}} \textbf{19} (2023), 064, 44~pages,
 \href{https://arxiv.org/abs/2201.11594}{arXiv:2201.11594}.

\bibitem{ghm}
Grassi A., Hatsuda Y., Mari\~no M., Topological strings from quantum mechanics,
 \href{https://doi.org/10.1007/s00023-016-0479-4}{\textit{Ann. Henri Poincar\'e}} \textbf{17} (2016), 3177--3235,
 \href{https://arxiv.org/abs/1410.3382}{arXiv:1410.3382}.

\bibitem{gkmw}
Grimm T.W., Klemm A., Mari\~no M., Weiss M., Direct integration of the
 topological string, \href{https://doi.org/10.1088/1126-6708/2007/08/058}{\textit{J.~High Energy Phys.}} \textbf{2007} (2007),
 no.~8, 058, 78~pages, \href{https://arxiv.org/abs/hep-th/0702187}{arXiv:hep-th/0702187}.

\bibitem{Gu:2023wum}
Gu J., Relations between {S}tokes constants of unrefined and
 {N}ekrasov--{S}hatashvili topological strings, \href{https://doi.org/10.1007/jhep05(2024)199}{\textit{J.~High Energy Phys.}}
 \textbf{2024} (2024), no.~5, 199, 29~pages, \href{https://arxiv.org/abs/2307.02079}{arXiv:2307.02079}.

\bibitem{Gu:2023mgf}
Gu J., Kashani-Poor A.K., Klemm A., Mari\~no M., Non-perturbative topological
 string theory on compact {C}alabi--{Y}au 3-folds, \href{https://doi.org/10.21468/scipostphys.16.3.079}{\textit{SciPost Phys.}}
 \textbf{16} (2024), 079, 84~pages, \href{https://arxiv.org/abs/2305.19916}{arXiv:2305.19916}.

\bibitem{gm-peacock}
Gu J., Mari\~no M., Peacock patterns and new integer invariants in topological
 string theory, \href{https://doi.org/10.21468/scipostphys.12.2.058}{\textit{SciPost Phys.}} \textbf{12} (2022), 058, 51~pages,
 \href{https://arxiv.org/abs/2104.07437}{arXiv:2104.07437}.

\bibitem{gm-multi}
Gu J., Mari\~no M., Exact multi-instantons in topological string theory,
 \href{https://doi.org/10.21468/scipostphys.15.4.179}{\textit{SciPost Phys.}} \textbf{15} (2023), 179, 36~pages,
 \href{https://arxiv.org/abs/2211.01403}{arXiv:2211.01403}.

\bibitem{gm-qp}
Gu J., Mari\~no M., On the resurgent structure of quantum periods,
  \href{https://doi.org/10.21468/SciPostPhys.15.1.035}{\textit{SciPost Phys.}} \textbf{15} (2023), 035, 40~pages,
 \href{https://arxiv.org/abs/2211.03871}{arXiv:2211.03871}.

\bibitem{hkr}
Haghighat B., Klemm A., Rauch M., Integrability of the holomorphic anomaly
 equations, \href{https://doi.org/10.1088/1126-6708/2008/10/097}{\textit{J.~High Energy Phys.}} \textbf{2008} (2008), no.~10, 097,
 37~pages, \href{https://arxiv.org/abs/0809.1674}{arXiv:0809.1674}.

\bibitem{ho}
Hatsuda Y., Okuyama K., Resummations and non-perturbative corrections,
 \href{https://doi.org/10.1007/JHEP09(2015)051}{\textit{J.~High Energy Phys.}} \textbf{2015} (2015), no.~9, 051, 28~pages,
 \href{https://arxiv.org/abs/1505.07460}{arXiv:1505.07460}.

\bibitem{hiv}
Hollowood T., Iqbal A., Vafa C., Matrix models, geometric engineering and
 elliptic genera, \href{https://doi.org/10.1088/1126-6708/2008/03/069}{\textit{J.~High Energy Phys.}} \textbf{2008} (2008), no.~3,
 069, 81~pages, \href{https://arxiv.org/abs/hep-th/0310272}{arXiv:hep-th/0310272}.

\bibitem{hkk}
Huang M., Kashani-Poor A.K., Klemm A., The~{$\Omega$} deformed {B}-model for
 rigid~{$\mathcal{N}=2$} theories,  \href{https://doi.org/10.1007/s00023-012-0192-x}{\textit{Ann. Henri Poincar\'e}} \textbf{14}
 (2013), 425--497, \href{https://arxiv.org/abs/1109.5728}{arXiv:1109.5728}.

\bibitem{hkatzk}
Huang M., Katz S., Klemm A., Towards refining the topological strings on
 compact {C}alabi--{Y}au 3-folds,  \href{https://doi.org/10.1007/jhep03(2021)266}{\textit{J.~High Energy Phys.}} \textbf{2021}
 (2021), no.~3, 266, 87~pages, \href{https://arxiv.org/abs/2010.02910}{arXiv:2010.02910}.

\bibitem{hk06}
Huang M., Klemm A., Holomorphic anomaly in gauge theories and matrix models,
  \href{https://doi.org/10.1088/1126-6708/2007/09/054}{\textit{J.~High Energy Phys.}} \textbf{2007} (2007), no.~9, 054, 33~pages,
 \href{https://arxiv.org/abs/hep-th/0605195}{arXiv:hep-th/0605195}.

\bibitem{hk}
Huang M., Klemm A., Direct integration for general~{$\Omega$} backgrounds,
  \href{https://doi.org/10.4310/ATMP.2012.v16.n3.a2}{\textit{Adv. Theor. Math. Phys.}} \textbf{16} (2012), 805--849,
 \href{https://arxiv.org/abs/1009.1126}{arXiv:1009.1126}.

\bibitem{ikv}
Iqbal A., Koz{\c{c}}az C., Vafa C., The refined topological vertex,
  \href{https://doi.org/10.1088/1126-6708/2009/10/069}{\textit{J.~High Energy Phys.}} \textbf{2009} (2009), no.~10, 069, 58~pages,
 \href{https://arxiv.org/abs/hep-th/0701156}{arXiv:hep-th/0701156}.

\bibitem{Iwaki:2023cek}
Iwaki K., Mari\~no M., Resurgent structure of the topological string and the
 first {P}ainlev\'e equation, \href{https://doi.org/10.3842/SIGMA.2024.028}{\textit{SIGMA}} \textbf{20} (2024), 028,
 21~pages, \href{https://arxiv.org/abs/2307.02080}{arXiv:2307.02080}.

\bibitem{Kidwai:2022kxx}
Kidwai O., Osuga K., Quantum curves from refined topological recursion: the
 genus~0 case, \href{https://doi.org/10.1016/j.aim.2023.109253}{\textit{Adv. Math.}} \textbf{432} (2023), 109253, 52~pages,
 \href{https://arxiv.org/abs/2204.12431}{arXiv:2204.12431}.

\bibitem{klemm}
Klemm A., The {B}-model approach to topological string theory on
 {C}alabi--{Y}au n-folds, in B-model {G}romov--{W}itten theory, \textit{Trends Math.},
 \href{https://doi.org/10.1007/978-3-319-94220-9_2}{Birkh\"auser}, Cham, 2018, 79--397.

\bibitem{kz}
Klemm A., Zaslow E., Local mirror symmetry at higher genus, in Winter {S}chool
 on {M}irror {S}ymmetry, {V}ector {B}undles and {L}agrangian {S}ubmanifolds,
 \textit{AMS/IP Stud. Adv. Math.}, Vol.~23,  \href{https://doi.org/10.1090/amsip/023/07}{American Mathematical Society},
 Providence, RI, 2001, 183--207, \href{https://arxiv.org/abs/hep-th/9906046}{arXiv:hep-th/9906046}.

\bibitem{KS:stability}
Kontsevich M., Soibelman Y., Stability structures, motivic
 {D}onaldson--{T}homas invariants and cluster transformations,
 \href{https://arxiv.org/abs/0811.2435}{arXiv:0811.2435}.

\bibitem{kw}
Krefl D., Walcher J., Extended holomorphic anomaly in gauge theory,
 \href{https://doi.org/10.1007/s11005-010-0432-2}{\textit{Lett. Math. Phys.}} \textbf{95} (2011), 67--88, \href{https://arxiv.org/abs/1007.0263}{arXiv:1007.0263}.

\bibitem{MR1327157}
LeBrun C., Fano manifolds, contact structures, and quaternionic geometry,
  \href{https://doi.org/10.1142/S0129167X95000146}{\textit{Internat.~J. Math.}} \textbf{6} (1995), 419--437,
 \href{https://arxiv.org/abs/dg-ga/9409001}{arXiv:dg-ga/9409001}.

\bibitem{mmnp}
Mari\~no M., Nonperturbative effects and nonperturbative definitions in matrix
 models and topological strings, \href{https://doi.org/10.1088/1126-6708/2008/12/114}{\textit{J.~High Energy Phys.}} \textbf{2008}
 (2008), no.~12, 114, 56~pages, \href{https://arxiv.org/abs/0805.3033}{arXiv:0805.3033}.

\bibitem{mm-open}
Mari\~no M., Open string amplitudes and large order behavior in topological
 string theory, \href{https://doi.org/10.1088/1126-6708/2008/03/060}{\textit{J.~High Energy Phys.}} \textbf{2008} (2008), no.~3,
 060, 34~pages, \href{https://arxiv.org/abs/hep-th/0612127}{arXiv:hep-th/0612127}.

\bibitem{mmbook}
Mari\~no M., Instantons and large~{$N$}. {A}n introduction to non-perturbative
 methods in quantum field theory, \href{https://doi.org/10.1017/CBO9781107705968}{Cambridge University Press}, Cambridge, 2015.

\bibitem{mm-s2019}
Mari\~no M., From resurgence to BPS states, {T}alk given at the conference {\it
 Strings 2019} (Brussels),
 \url{https://member.ipmu.jp/yuji.tachikawa/stringsmirrors/2019/2_M_Marino.pdf}.

\bibitem{msw}
Mari\~no M., Schiappa R., Weiss M., Nonperturbative effects and the large-order
 behavior of matrix models and topological strings, \href{https://doi.org/10.4310/CNTP.2008.v2.n2.a3}{\textit{Commun. Number
 Theory Phys.}} \textbf{2} (2008), 349--419, \href{https://arxiv.org/abs/0711.1954}{arXiv:0711.1954}.

\bibitem{ms}
Mari\~no M., Schwick M., Non-perturbative real topological strings,
 \href{https://arxiv.org/abs/2309.12046}{arXiv:2309.12046}.

\bibitem{gw-dt}
Maulik D., Nekrasov N., Okounkov A., Pandharipande R., Gromov--{W}itten theory
 and {D}onaldson--{T}homas theory.~{I}, \href{https://doi.org/10.1112/S0010437X06002302}{\textit{Compos. Math.}} \textbf{142}
 (2006), 1263--1285, \href{https://arxiv.org/abs/math.AG/0312059}{arXiv:math.AG/0312059}.

\bibitem{gw-dt2}
Maulik D., Nekrasov N., Okounkov A., Pandharipande R., Gromov--{W}itten theory
 and {D}onaldson--{T}homas theory.~{II}, \href{https://doi.org/10.1112/S0010437X06002314}{\textit{Compos. Math.}} \textbf{142}
 (2006), 1286--1304, \href{https://arxiv.org/abs/math.AG/0406092}{arXiv:math.AG/0406092}.

\bibitem{Mozgovoy:2020has}
Mozgovoy S., Pioline B., Attractor invariants, brane tilings and crystals, \textit{Ann. Inst. Fourier (Grenoble)}, to appear,
 \href{https://arxiv.org/abs/2012.14358}{arXiv:2012.14358}.

\bibitem{Neitzke:2011za}
Neitzke A., On a hyperholomorphic line bundle over the {C}oulomb branch,
 \href{https://arxiv.org/abs/1110.1619}{arXiv:1110.1619}.

\bibitem{n}
Nekrasov N., Seiberg--{W}itten prepotential from instanton counting,
 \href{https://doi.org/10.4310/ATMP.2003.v7.n5.a4}{\textit{Adv. Theor. Math. Phys.}} \textbf{7} (2003), 831--864,
 \href{https://arxiv.org/abs/hep-th/0206161}{arXiv:hep-th/0206161}.

\bibitem{no2}
Nekrasov N., Okounkov A., Seiberg--{W}itten theory and random partitions, in
 The {U}nity of {M}athematics, \textit{Progr. Math.}, Vol. 244, \href{https://doi.org/10.1007/0-8176-4467-9_15}{Birkh\"auser},
 Boston, MA, 2006, 525--596, \href{https://arxiv.org/abs/hep-th/0306238}{arXiv:hep-th/0306238}.

\bibitem{no}
Nekrasov N., Okounkov A., Membranes and sheaves, \href{https://doi.org/10.14231/AG-2016-015}{\textit{Algebr. Geom.}}
 \textbf{3} (2016), 320--369, \href{https://arxiv.org/abs/1404.2323}{arXiv:1404.2323}.

\bibitem{ns}
Nekrasov N., Shatashvili S., Quantization of integrable systems and four
 dimensional gauge theories, in X{VI}th {I}nternational {C}ongress on
 {M}athematical {P}hysics, \href{https://doi.org/10.1142/9789814304634_0015}{World Scientific Publishing}, Hackensack, NJ, 2010,
 265--289, \href{https://arxiv.org/abs/0908.4052}{arXiv:0908.4052}.

\bibitem{Nekrasov:2010ka}
Nekrasov N., Witten E., The omega deformation, branes, integrability and
 {L}iouville theory, \href{https://doi.org/10.1007/JHEP09(2010)092}{\textit{J.~High Energy Phys.}} \textbf{2010} (2010),
 no.~9, 092, 82~pages, \href{https://arxiv.org/abs/1002.0888}{arXiv:1002.0888}.

\bibitem{ps09}
Pasquetti S., Schiappa R., Borel and {S}tokes nonperturbative phenomena in
 topological string theory and {$c=1$} matrix models, \href{https://doi.org/10.1007/s00023-010-0044-5}{\textit{Ann. Henri
 Poincar\'e}} \textbf{11} (2010), 351--431, \href{https://arxiv.org/abs/0907.4082}{arXiv:0907.4082}.

\bibitem{Shenker:1990uf}
Shenker S.H., The strength of nonperturbative effects in string theory, in
 Random {S}urfaces and {Q}uantum {G}ravity, \textit{NATO Adv. Sci. Inst.
 Ser.~B: Phys.}, Vol. 262, \href{https://doi.org/10.1007/978-1-4615-3772-4_12}{Plenum}, New York, 1991, 191--200.

\bibitem{MR1096180}
Swann A., Hyper{K}\"ahler and quaternionic {K}\"ahler geometry, \href{https://doi.org/10.1007/BF01446581}{\textit{Math.
 Ann.}} \textbf{289} (1991), 421--450.

\bibitem{yy}
Yamaguchi S., Yau S.-T., Topological string partition functions as polynomials,
  \href{https://doi.org/10.1088/1126-6708/2004/07/047}{\textit{J.~High Energy Phys.}} \textbf{2004} (2004), no.~7, 047, 20~pages,
 \href{https://arxiv.org/abs/hep-th/0406078}{arXiv:hep-th/0406078}.

\end{thebibliography}
\end{document}